\documentclass[aps,showkeys,showpacs,notitlepage,superscriptaddress]{revtex4-1}
\pdfoutput=1
\usepackage{amsmath}
\usepackage{amssymb}
\usepackage{amsfonts}
\usepackage{graphicx}
\usepackage{hyperref}
\usepackage{color}
\usepackage{multirow}
\usepackage{mathtools}
\usepackage{booktabs}
\usepackage{amsthm}
\usepackage{algorithm}
\usepackage{algpseudocode}
\usepackage{mathtools}
\usepackage{adjustbox}
\usepackage{xcolor}
\usepackage{url}
\usepackage{tikz-cd}
\usepackage{ulem}
\hypersetup{pdftitle=Persistent homology of unweighted networks via discrete Morse theory}

\def\rank{\operatorname{rank}}

\def\id{\operatorname{id}}

\def\maK{\mathcal{K}}

\def\maG{\mathcal{G}}
\def\maF{\mathcal{F}}

\def\R{\mathbb R}

\def\Z{\mathbb Z}
\def\N{\mathbb N}

\def\maF{{\mathcal F}}

\begin{document}
\title{Persistent homology of unweighted complex networks via discrete Morse theory}

\author{Harish Kannan}
\affiliation{The Institute of Mathematical Sciences (IMSc), Homi Bhabha National Institute (HBNI), Chennai 600113 India}
\author{Emil Saucan}
\affiliation{Department of Applied Mathematics, ORT Braude College, Karmiel 2161002 Israel}
\affiliation{Department of Electrical Engineering, Technion, Israel Institute of Technology, Haifa 3200003 Israel}
\author{Indrava Roy}
\email{Correspondence to: indrava@imsc.res.in}
\affiliation{The Institute of Mathematical Sciences (IMSc), Homi Bhabha National Institute (HBNI), Chennai 600113 India}
\author{Areejit Samal}
\email{Correspondence to: asamal@imsc.res.in}
\affiliation{The Institute of Mathematical Sciences (IMSc), Homi Bhabha National Institute (HBNI), Chennai 600113 India}
\affiliation{Max Planck Institute for Mathematics in the Sciences, Leipzig 04103 Germany}

\begin{abstract}
Topological data analysis can reveal higher-order structure beyond pairwise connections between vertices in complex networks. We present a new method based on discrete Morse theory to study topological properties of unweighted and undirected networks using persistent homology. Leveraging on the features of discrete Morse theory, our method not only captures the topology of the clique complex of such graphs via the concept of critical simplices, but also achieves close to the theoretical minimum number of critical simplices in several analyzed model and real networks. This leads to a reduced filtration scheme based on the subsequence of the corresponding critical weights, thereby leading to a significant increase in computational efficiency. We have employed our filtration scheme to explore the persistent homology of several model and real-world networks. In particular, we show that our method can detect differences in the higher-order structure of networks, and the corresponding persistence diagrams can be used to distinguish between different model networks. In summary, our method based on discrete Morse theory further increases the applicability of persistent homology to investigate the global topology of complex networks.
\end{abstract}

\maketitle

\section*{Introduction}

In recent years, the field of topological data analysis (TDA) has rapidly grown to provide a set of powerful tools to analyze various important features of data \cite{Carlsson2009}. In this context, persistent homology has played a key role in bringing TDA to the fore of modern data analysis. It not only gives a way to visualize data efficiently, but also to extract relevant information from both structured and unstructured datasets. This crucial aspect has been used effectively in various applications from astrophysics (e.g., determination of inter-galactic filament structures \cite{Pranav2016}) to imaging analysis (e.g., feature detection in 3D gray-scale images \cite{Gunther2011}) to biology (e.g., detection of breast cancer type with high survival rates \cite{Nicolau2011}). Informally, the essence of the theory is its power to extract the \textit{shape of data}, as well as infer higher-order correlations between various parts of the data at hand which are missed by other classical techniques \cite{Carlsson2009}. The basic mathematical theory used in this subject is that of algebraic topology, and in particular the study of homology, developed by the French mathematician Henri Poincar\'{e} at the turn of the 20th century. The origins of persistent homology lie in the ideas of Morse theory \cite{Morse1934}, which gives a powerful tool to detect the topological features of a given space through the computation of homology using real-valued functions on the space. We refer the reader to the survey article \cite{Edelsbrunner2008} for further details.

On the other hand, the discretized version of Morse theory developed by Robin Forman \cite{Forman1995,Forman1998,Forman2002}, gives a way to characterize the homology group of a simplicial complex in terms of a real-valued function with certain properties, known as a discrete Morse function. Examples of such simplicial complexes associated with discrete spaces are the Vietoris-Rips complex corresponding to a discrete metric space, or the clique complex of a graph. Forman \cite{Forman1998,Forman2002} showed that given such a function, the so-called critical simplices completely determine the Euler characteristic of the space, which is a fundamental topological invariant.

The study of complex networks in the last few decades has also significantly raised our ability to understand various kinds of interactions arising in both natural and artificial realms \cite{Watts1998,Barabasi1999,Albert2002,Newman2010}. Understanding how different parts of networks behave and influence each other is therefore an important problem \cite{Watts1998,Barabasi1999,Albert2002,Newman2010}. However, for large networks, detecting higher-order structures remains a difficult task \cite{Bianconi2015}. Moreover, recent studies \cite{Kartun-Giles2019,Iacopini2019,Ritchie2017} indicate that these higher-order correlations are not captured by usual network measures such as clustering coefficients. While a graph representation captures binary relationships among vertices of a network, simplicial complexes also reflect higher-order relationships in a complex network \cite{De2007,Horak2009,Petri2013,Petri2014,Wu2015,Sizemore2016,Courtney2017,Courtney2018,Kartun-Giles2019,Iacopini2019}. In this context, persistent homology has been employed to explore the topological properties of complex networks \cite{De2007,Horak2009,Lee2012,Petri2013,Petri2014,Sizemore2016}. In this work, we present a systematic method to study the persistent homology of unweighted and undirected graphs or networks.

Previous work has investigated the persistent homology of weighted and undirected networks by creating a filtration of the clique complexes corresponding to threshold graphs obtained via decreasing sequence of edge weights \cite{Petri2013,Sizemore2016}. However, the lack of edge weights in unweighted networks does not permit a filtration based on threshold graphs \cite{Petri2013,Sizemore2016}. Thus for unweighted networks, Horak \textit{et al.} \cite{Horak2009} propose a filtration scheme based on the dimension of the simplices in the clique complex corresponding to the unweighted network. Horak \textit{et al.} \cite{Horak2009} do not assign weights to vertices, edges or higher-dimensional simplices in the clique complex corresponding to an
unweighted graph. An unexplored filtration scheme involves transforming an unweighted network into a weighted network by assigning edge weights based on
some network property, such as edge betweenness centrality \cite{Freeman1977,Girvan2002} or discrete edge curvature \cite{Sreejith2016,Samal2018}, and then employing the filtration scheme based on threshold graphs \cite{Petri2013,Sizemore2016}. As an alternative, we here use discrete Morse theory \cite{Forman1995,Forman1998,Forman2002} to create a filtration scheme for unweighted networks by assigning weights to vertices, edges, triangles and higher-dimensional simplices in the clique complex of the graph. In our method, the weight of a simplex is chosen such that it reflects the degree of the vertices which constitute the simplex while simultaneously satisfying the conditions for the weighing function to be a discrete Morse function. Moreover, as explained in the Results section, an equally important intuition behind the choice of these weights is based on the goal of reducing the number of so-called critical simplices.

In the context of TDA, classical Morse theory, which involves smooth functions defined on topological spaces that admit a smooth structure has been used to compute persistent homology, e.g. in statistical topology \cite{Bubenik2009}, astrophysics \cite{Pranav2016}. Since the clique complex of a weighted or unweighted graph does not permit a smooth structure in general, applying classical Morse theory is not possible. However, discrete Morse theory \cite{Forman1995,Forman1998,Forman2002} provides an efficient way of computing persistent homology. A discrete Morse function not only captures higher-order topological information of the underlying space, a ``preprocessing'' with respect to a suitable discrete Morse function leads to
significant simplification of the topological structure. This makes computation of persistent homology groups or homology groups of filtered complexes much more efficient, see e.g. \cite{Mischaikow2013}. This is especially useful for large datasets where computationally efficient methods are key to compute their persistent homology.

We have applied discrete Morse theory to compute persistent homology of unweighted simple graphs. This is done by using the values given by the discrete
Morse function to pass from an unweighted graph to a weighted simplicial complex (Figure \ref{fig1}). This transformation automatically produces a filtration that is needed for the computation of persistent homology, through the so-called level subcomplexes associated with weights of critical simplices (See Theory section and Figure \ref{fig2}). Moreover, this filtration is consistent with the topology of the underlying space and reveals finer topological features than the dimensional filtration scheme used in Horak \textit{et al.}\cite{Horak2009} The combination of these techniques have been used \cite{Gunther2011, Delgado2014} with applications for image processing. However, to the best of our knowledge, this method has not been used for studying persistent homology in unweighted complex networks to date. Discrete Morse theory gives a theoretical lower bound on the number of critical simplices which can be attained by an optimal choice of the function on a simplicial complex. Interestingly, our method achieves close to the theoretical minimum number of critical simplices in several model and real networks analyzed here (See Results section). Furthermore, our algorithm for computing the discrete Morse function is easy to implement for complex networks.

Our results underline the potence of persistent homology to detect inherent topological features of networks which are not directly captured by homology
alone. For instance, the $p$-Betti numbers of the clique complexes corresponding to small-world \cite{Watts1998} and scale-free \cite{Barabasi1999} networks with similar size and average degree, respectively, are of comparable magnitude and thus, homology reveals no deep insight into the differences between the topological features of these two model networks. On the other hand, our observations on the persistent homology of these two networks indicate a clear demarkation with respect to the evolution of topological features in the clique complexes corresponding to these model networks during the filtration process. This dissimilarity in the evolution of topological characteristics that resonates across dimensions and the average degree of the underlying network, indicates an inherent disparity in the persistent homology of small-world and scale-free graphs. In addition to unravelling higher-order relationships in networks, this ability to capture inherent topological differences between two dissimilar networks thus motivates the application of our methods to study the persistent homology of real-world networks.

The remainder of the paper is organized as follows. We begin with a \textbf{Theory} section which gives a brief overview of concepts in persistent homology and discrete Morse theory. We then proceed to describe the model networks and real-world networks that have been studied in this work in the \textbf{Network datasets} section. In the subsequent section on \textbf{Results and Discussion}, we present our algorithm to construct a discrete Morse function on a simplical complex associated with a network. In the same section, we present our results for model networks and real-world networks. The final section on \textbf{Conclusions} gives a summary and outlook of our findings. In Supplementary Information (SI) \textbf{Appendix}, we give a brief review of the mathematical theory of homology groups of a simplicial complex. In SI Appendix, we also provide a rigorous proof of concept for our algorithm to construct a discrete Morse function. We then follow up with two algorithms both of which illustrate key procedures that are essential to construct the filtration of a simplical complex associated with the investigated networks. In SI Appendix, time complexity for our algorithms as well as some theoretical results on stability of persistent homology and persistence diagrams are also given.


\section*{Theory}

\subsection*{Graphs and Simplicial Complexes}

Consider a finite simple graph $G(\mathcal{V},\mathcal{E})$ having vertex set $\mathcal{V}=\{v_0,v_1,\cdots,v_n\}$ and the edge set $\mathcal{E}$. Note that a simple graph does not contain self-loops or multi-edges \cite{Bollobas1998}. Such a simple graph $G$ can be viewed as a \textit{clique} complex $K$ \cite{Zomorodian2005}. A clique simplicial complex $K$ is a collection of simplices where a $p$-dimensional simplex (or $p$-simplex) in $K$ is a set of
$p+1$ vertices that form a complete subgraph. In other words, vertices correspond to $0$-simplices, edges to $1$-simplices, and triangles to
$2$-simplices in the clique complex of a graph. Note that the dimension $p$ of simplices contained in $K$ is restricted to the range 0 to $(|\mathcal{V}| - 1)$ in the graph $G$. The dimension $d$ of the clique complex $K$ is given by the maximum dimension of its constituent simplices. A \textit{face} $\gamma$ of a $p$-simplex $\alpha$ is a subset of $\alpha$ with cardinality less than $p+1$. Note that by definition, a face $\gamma$ of a $p$-simplex $\alpha$ is a $l$-simplex where $0 \le l < p$ and this relationship is denoted as $\gamma^{l}<\alpha^{p}$. Formally, the clique complex $K$ corresponding to the simple graph $G$ satisfies the following condition which defines an abstract simplicial complex, namely, $K$ is a collection of non-empty finite sets or simplices such that if $\alpha$ is an element (simplex) of $K$ then so is every non-empty subset of $\alpha$. For additional details, the interested reader is referred to standard text in algebraic topology \cite{Munkres2018}.

\begin{figure*}
\includegraphics[width=.97\columnwidth]{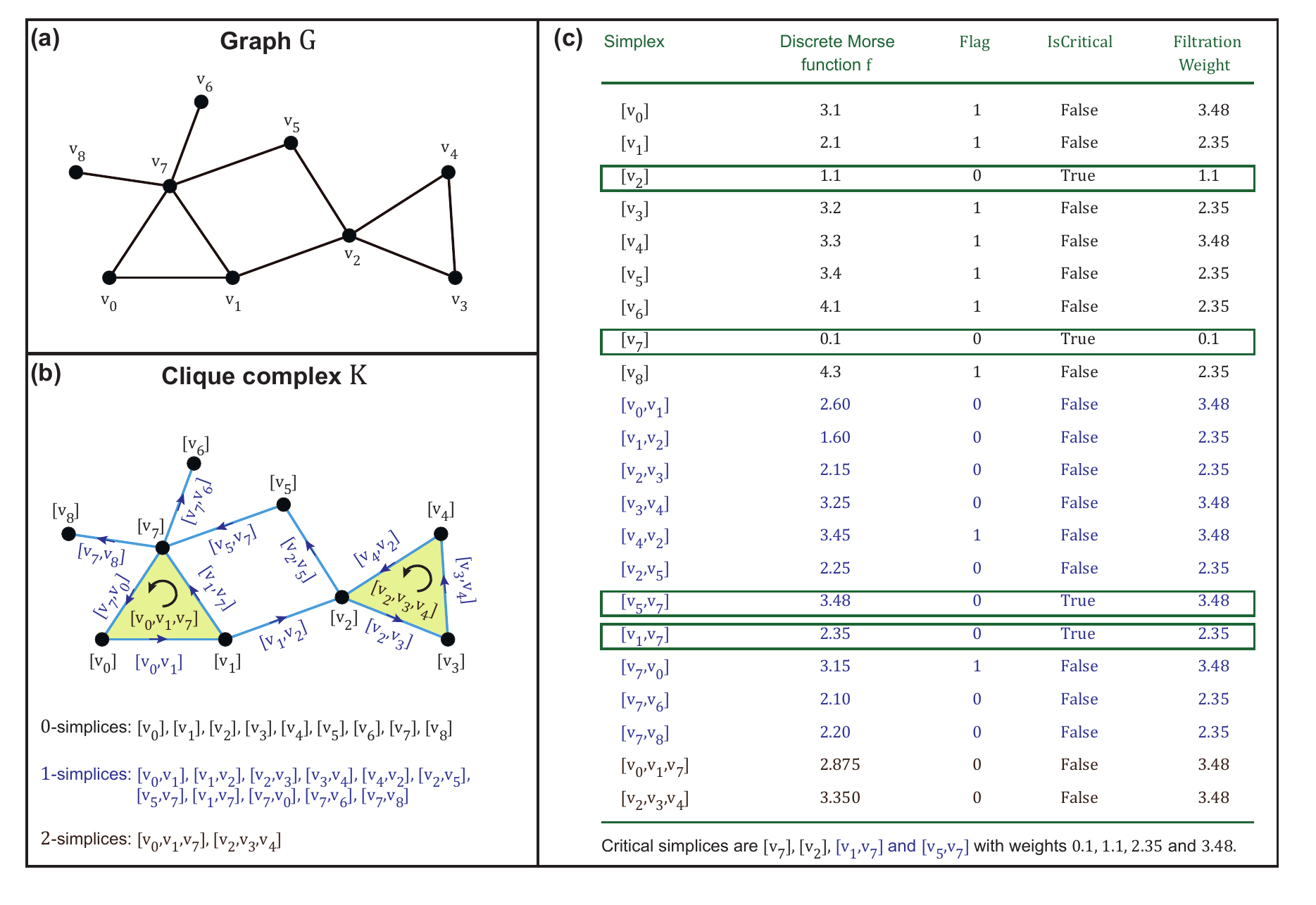}
\caption{An illustration of the construction of a discrete Morse function $f$ on a clique complex $K$ corresponding to an unweighted and undirected graph $G$ using our algorithm. (a) A simple example of an unweighted and undirected graph $G$ containing 9 vertices and 11 edges. (b) The clique simplicial complex $K$ corresponding to the simple graph $G$ shown in (a). The clique complex $K$ consists of 9 vertices or $0$-simplices, 11 edges or  $1$-simplices and 2 triangles or $2$-simplices. The figure also displays the orientation of the $1$- and $2$-simplices using arrows. (c) Generation of a discrete Morse function $f$ on the clique complex $K$ shown in (b) using our algorithm. The figure lists the state of the \texttt{Flag} variable in algorithm 1 and \texttt{IsCritical} variable in algorithm 2 (See SI Appendix) for each simplex in $K$. In this example, the clique complex has 4 critical simplices and their respective critical weights correspond to the filtration steps. The figure also lists the \texttt{FiltrationWeight} for each simplex in $K$ obtained using algorithm 3 (See SI Appendix).}
\label{fig1}
\end{figure*}

Figure \ref{fig1} displays an example of the correspondence between a simple graph and its clique complex. The ordering of the vertex set $\{v_0,v_1,v_2,
\cdots,v_p\}$ of a $p$-simplex $\alpha$ determines its \textit{orientation}. Moreover, two orderings of the vertex set of $\alpha$ are considered to be
equivalent if and only if they differ by an even permutation. If the dimension of a $p$-simplex is greater than 1, then all possible orderings of its vertex set fall under two equivalence classes, with each class being assigned an \textit{orientation} \cite{Munkres2018}. An exception is the $0$-simplex with one vertex which has exactly one equivalence class and orientation. An oriented $p$-simplex $\alpha$ specifies the orientation of its $p+1$ vertices and is represented by $[v_0,v_1, v_2,\cdots,v_p]$ \cite{Munkres2018}. In figure \ref{fig1}, the oriented $2$-simplices $[v_2,v_3,v_4]$ and $[v_2,v_4,v_3]$ have opposite orientations, i.e., $[v_2,v_3,v_4] = -[v_2,v_4,v_3]$.


\subsection*{Persistent homology of a simplicial complex}

In SI Appendix, we give a brief review of the mathematical theory of homology groups of a simplicial complex. In particular, we define $p$-chain group, $p$-boundary operator, $p$-boundary, $p$-cycle, $p$-hole, $p$-homology group and $p$-Betti number.

A subset $K^i$ of a simplicial complex $K$ is called a subcomplex of $K$ if $K^i$ by itself is an abstract simplicial complex. Then a filtration of a
simplicial complex $K$ is defined as a nested sequence of subcomplexes $K^i$ of $K$ where:
\begin{equation}
\emptyset \subseteq K^0 \subseteq K^1 \subseteq ... \subseteq K^q = K
\end{equation}
Note that each subcomplex $K^i$ has an associated index $i$ in the filtration. Moreover, each subcomplex $K^i$ in the filtration has corresponding $p$-chain complexes $C_p^i$, $p$-boundary operators $\partial_p^i$, $p$-boundaries $B_p^i$ and $p$-cycles $Z_p^i$.

The $j$-persistent $p$-homology group of $K^i$ denoted as $H_p^{i,j}$ is defined
as:
\begin{equation}
H_p^{i,j} = Z_p^i / ( B_p^{i+j} \cap Z_p^i).
\end{equation}
In the above equation, $B_p^{i+j}$ is the subgroup of $C_p^{i+j}$ which constitutes the $p$-boundaries of the subcomplex $K^{i+j}$. The $j$-persistent
$p$-Betti number of $K^i$ denoted as $\beta_p^{i,j}$ is defined as:
\begin{equation}
\beta_p^{i,j} = \text{dim}(H_p^{i,j}).
\end{equation}
An intuitive explanation of the above definitions of the $j$-persistent $p$-homology group and the corresponding Betti number is as follows. A $p$-hole
of the subcomplex $K^i$ can potentially become the boundary of a $(p+1)$-chain of a later subcomplex $K^{i+j}$ with $j>0$, and thus, no longer constitute a $p$-hole of $K^{i+j}$. The $j$-persistent $p$-Betti number of $K^i$ represents the number of $p$-holes at the filtration index $i$ that persist at the
filtration index $j+i$. Therefore, each $p$-hole that appears across the filtration has a unique index that corresponds to its \textit{birth} and
\textit{death}, and the persistence of such a $p$-hole can thus be characterized by its corresponding birth and death indices. Studying persistent homology allows us to quantify the longevity of such $p$-holes during filtration, and thus, measures the importance of these topological features which appear and disappear across the filtration.


\begin{figure*}
\includegraphics[width=.81\columnwidth]{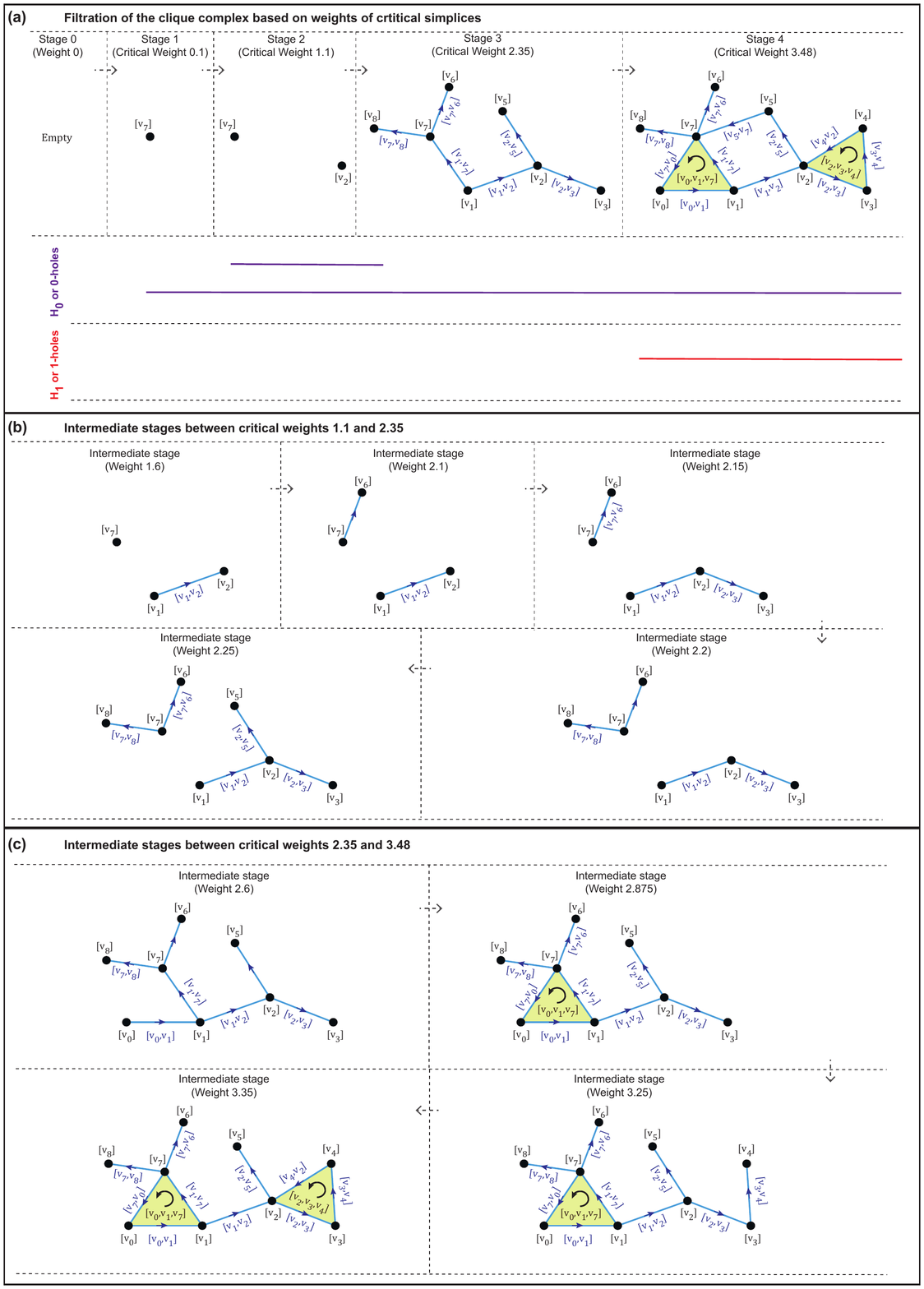}
\caption{Filtration based on the entire sequence of weights satisfying discrete Morse function is equivalent to filtration based only on the subsequence of critical weights in terms of persistent homology. (a) Filtration of the network shown in Figure \ref{fig1} based on weights of the 4 critical simplices. There is a $0$-hole (or connected component) that persists across the 4 stages of the filtration while another 0-hole is born at stage 2 on addition of critical vertex $v_2$ but dies at the stage 3 which corresponds to the weight of the critical edge $[v_1,v_7]$. Moreover, a $1$-hole is born at the stage 4 on addition of the critical edge $[v_5,v_7]$. (b) Five intermediate stages during the filtration between critical weights 1.1 (stage 2) and 2.35 (stage 3). (c) Four intermediate stages during the filtration between critical weights 2.35 (stage 3) and 3.48 (stage 4). It is seen that the homology of the clique complex remains unchanged during the intermediate stages of the filtration whereby the birth and death of holes occur only at stages which correspond to critical weights.}
\label{fig2}
\end{figure*}

\subsection*{Discrete Morse Theory}

Recalling from the preceeding section, to study the persistent homology corresponding to the clique complex $K$ of a simple graph $G$, the primary
requirement is a filtration of $K$. We here present a systematic method to study the persistent homology of unweighted and undirected networks by utilizing a refined filtration of the clique complex $K$ based on discrete Morse theory \cite{Forman1995,Forman1998,Forman2002}. It is important to note that the order in which the simplices of the clique complex $K$ are added during the filtration affects the evolution of topological features which are observed by studying the persistent homology. Our proposed scheme which is based on discrete Morse theory developed by Forman \cite{Forman1995,Forman1998,Forman2002} tackles this by assigning weights to $0$-simplices (vertices), $1$-simplices (edges), $2$-simplices (triangles) and higher-dimensional simplices appearing in the clique complex corresponding to an unweighted and undirected network. Assigning weights to higher-dimensional simplices captures important higher-order correlations in addition to edges or $1$-simplices. Moreover, we leverage the following important features of the framework of discrete Morse theory in our new scheme. Firstly, the framework enables assignment of weights to
$p$-simplices which are concordant with weights of $(p-1)$-simplices. Secondly, it captures the topology of a simplicial complex via the concept of
\textit{critical} simplices described below. Most importantly, the framework provides a natural way to create a filtration scheme to study persistent
homology based upon the weights of the aforementioned \textit{critical} simplices as will be described below.

We next provide the fundamental definitions in discrete Morse theory
\cite{Forman1998,Forman2002}.
We remark that a $p$-dimensional simplex $\alpha$ in a simplicial complex $K$ is
denoted by $\alpha^{p} \in K$.
Also, if a $p$-simplex $\alpha^{p}$ in $K$ is a face of a $(p+1)$-simplex $\beta^{p+1}$ in $K$ then this is represented as $\alpha^p < \beta^{p+1}$ in the sequel. Given a function $f : K \rightarrow \mathbb{R}$, for each simplex $\alpha^{p} \in K$, two sets $U_{\alpha}^f$ and $V_{\alpha}^f$ are defined as follows:
\begin{eqnarray}
\label{set-u}
U_{\alpha}^f = \{ \beta^{p+1} \: | \: \alpha^p < \beta^{p+1} \text{ and }
f(\beta) \leq f(\alpha) \}\\
\label{set-v}
V_{\alpha}^f = \{ \gamma^{p-1} \: | \: \gamma^{p-1} < \alpha^p \text{ and }
f(\alpha) \leq f(\gamma) \}
\end{eqnarray}
Simply stated, the set $U_{\alpha}^f$ contains any $(p+1)$-simplex $\beta^{p+1}$ of which $\alpha^p$ is a face and the function value on $\beta$ is less than or equal to the function value on $\alpha$. The set $V_{\alpha}^f$ contains any $(p-1)$-simplex $\gamma^{p-1}$ which is a face of $\alpha^p$ and the function value on $\alpha$ is less than or equal to the function value on $\gamma$. A function $f : K \rightarrow \mathbb{R}$ is a discrete Morse function \cite{Forman1998,Forman2002} if and only if for each simplex $\alpha^{p} \in K$:
\begin{equation}
\label{discmor}
| \: U_{\alpha}^f \: | \leq 1 \text{ and } | \: V_{\alpha}^f \: | \leq 1.
\end{equation}
Given a discrete Morse function $f$ on the simplicial complex $K$, a simplex $\alpha^{p} \in K$ is critical \cite{Forman1998,Forman2002} if and only if:
\begin{equation}
\label{critsimp}
| \: U_{\alpha}^f \: | = 0 \text{ and } | \: V_{\alpha}^f \: | = 0.
\end{equation}
Simply stated, a $p$-simplex $\alpha^{p} \in K$ is critical if the following conditions are simultaneously satisfied. The first condition being that if
$\beta^{p+1}$ is any $(p+1)$-simplex in $K$ of which $\alpha^p$ is a face, then $f(\alpha) < f(\beta)$. The second condition being that if $\gamma^{p-1}$ is any $(p-1)$-simplex in $K$ which is a face of $\alpha^p$ then $f(\alpha) > f(\gamma)$. The concept of critical simplices in discrete Morse theory is in spirit a discrete analogue to the concept of critical points in classical Morse theory wherein the critical points corresponding to a smooth real valued
function on $X \subseteq \mathbb{R}^d$ are the points where the gradient of the function vanishes.

We remark that once a discrete Morse function $f$ on a simplicial complex $K$ is fixed, the sets $U_{\alpha}^f$ and $V_{\alpha}^f$ are denoted by $U_{\alpha}$ and $V_{\alpha}$, respectively, to simplify the notation. A simple example for a discrete Morse function on a simplicial complex $K$ is the dimension function used in Horak {\it et al} \cite{Horak2009}. The value of the dimension function on a given simplex $\alpha$ is the dimension of the simplex $\alpha$. By a direct consequence of the definitions presented above in this section, for every simplex $\alpha \in K$, the sets $U_{\alpha}$ and $V_{\alpha}$ corresponding to the dimension function are empty. Thus, the dimension function is indeed a discrete Morse function and every simplex in $K$ is critical. In the results section, we present our new scheme and algorithm 1 to assign a discrete Morse function $f$ to a clique complex $K$ of an unweighted graph $G$.

We next describe the filtration of the clique simplicial complex $K$ based on the discrete Morse function $f$. Given a discrete Morse function $f$ on a
simplicial complex $K$ and a real number $r$, a level subcomplex $K(r)$ is defined \cite{Forman1998,Forman2002} as follows:
\begin{equation}
K(r) = \bigcup_{f(\beta) \leq r} \bigcup_{\alpha \leq \beta} \alpha
\end{equation}
Simply stated, $K(r)$ contains all simplices $\beta$ in $K$ with the value of the discrete Morse function or assigned weight $f(\beta) \leq r$ along with any face $\alpha$ of $\beta$. Note that a face $\alpha$ of $\beta$ is included in $K(r)$ even if the discrete Morse function or assigned weight to a face $\alpha$ is greater than $r$.

Let $\{ f(\sigma) \}_{\sigma \in K}$ denote the entire set of values assigned to simplices in $K$ using the discrete morse function $f$. Then, let $\{w_k
\}_{k=0,\cdots,n} $ denote the finite increasing sequence of the unique values in the set $\{ f(\sigma) \}_{\sigma \in K}$ associated with the finite
simplicial complex considered here. We now have a sequence of inclusions of level subcomplexes corresponding to this increasing sequence $\{ w_k \}$ as
follows:
\begin{equation}
\label{filtwt}
\emptyset \subseteq K(w_0) \subseteq  K(w_1)  \subseteq  \cdots \subseteq
K(w_{n-1}) \subseteq  K(w_n) = K.
\end{equation}
This nested sequence gives a filtration of the simplicial complex $K$ which enables the study of persistent homology in the context of unweighted networks.

According to \textit{Lemma 2.6} by Forman \cite{Forman2002}, if there are no critical simplices $\alpha$ with $f(\alpha) \in (a,b],$ then $K(b)$ is
\textit{homotopy equivalent} to $K(a)$.

The implications of this Lemma are as follows. Let $\{ f(\sigma_c) \}$ denote the set of values assigned to critical simplices $\sigma_c$ in $K$ by the
discrete Morse function $f$ and $\{ w_{c_k} \}_{k=0,\cdots,m}$ denote the increasing sequence of the unique values in $\{ f(\sigma_{c}) \}$. We refer to
the function values $\{ f(\sigma_c) \}$ assigned to the critical simplices $\sigma_c$ in $K$ as \textit{critical weights}. Note that the set $\{
f(\sigma_c) \}$ defined for critical simplices is a subset of $\{ f(\sigma) \}$ defined for all simplices in $K$ and the increasing sequence $\{ w_{c_k}
\}_{k=0,\cdots,m}$ is a subsequence of $\{w_k \}_{k=0,\cdots,n}$ with $m \leq n$. The above definition implies that there are no critical simplices $\alpha$ with $f(\alpha) \in (w_{c_i}, w_{c_{i+1}})$. As homology is invariant under homotopy equivalence, Forman's Lemma 2.6 gives us that for any $x$ and $y$ belonging to the real number interval $(w_{c_i}, w_{c_{i+1}})$, the homology groups of $K(x)$ and $K(y)$ are isomorphic. Thus, in order to observe the changes in homology as the filtration proceeds, it suffices to study the persistent homology of a filtration which corresponds to the subsequence $\{
w_{c_k} \}_{k=0,\cdots,m}$ of $\{ w_k \}_{k=0,\cdots,n}$, where $m \leq n$, and this results in a potential decrease in the required number of filtration
steps. The new filtration sequence can be represented as:
\begin{equation}
\label{filtcrit}
\emptyset\subseteq  K(w_{c_0}) \subseteq  K(w_{c_1}) \subseteq \cdots \subseteq
K(w_{c_{m-1}}) \subseteq  K(w_{c_m}) \subseteq K.
\end{equation}
Note that each simplex $\alpha$ in the clique complex $K$ is first introduced as part of certain level subcomplex $K(w_{c_i})$ in the above nested filtration sequence. Therefore, each simplex $\alpha$ in $K$ can be associated with a unique weight $w_{c_i}$ referred to as the filtration weight of $\alpha$. In SI Appendix, we present algorithm 2 and algorithm 3 which depict the procedure to compute the filtration weights of simplices in the clique complex $K$ of a graph $G$. In SI Appendix, we also give a sufficient condition for two discrete Morse functions to induce the same filtration and thus the persistent homology groups.

Using an example network in figure \ref{fig2}, we also show that the persistent homology observed using the filtration based on the entire sequence of weights satisfying discrete Morse function is equivalent to that observed using the filtration based on the subsequence of critical weights.

Let $m_p$ represent the number of critical $p$-simplices in a simplicial complex $K$ and let $\beta_p$ denote the $p$-Betti number of $K$. Then \textit{Theorem 2.11} by Forman \cite{Forman2002} can be stated as follows.
\begin{itemize}
\item[(i)] For each  $p = 0, 1, 2, \cdots, d$ (where $d$ is the dimension of $K$), $m_p \geq \beta_p$.
\item[(ii)] $m_0 - m_1 + m_2 - \cdots + (-1)^dm_d = \beta_0 - \beta_1 + \beta_2 - \cdots + (-1)^d\beta_d$.
\end{itemize}
In other words, the above theorem gives a lower bound of the number of $p$-critical simplices $m_p$ for each dimension $p$ as the $p$-Betti number
$\beta_p$ of $K$. In results section, we present our algorithm 1 to assign weights satisfying discrete Morse function to simplices in the clique complex
$K$ of a graph $G$. Our choice of the function in algorithm 1 to assign weights to simplices in the clique complex $K$ tries to minimize the number of critical simplices (which has a lower bound given by Forman's Theorem 2.11 \cite{Forman2002}), and thus, reduces the number of filtration steps required to compute the persistent homology without loss of information. In the results section, we will show that our algorithm achieves near-optimal number of
critical weights in clique complexes corresponding to many model and real networks analyzed here.


\subsection*{Comparing Persistence diagrams}

Given a discrete Morse function $f$ and its associated filtration $\{K(w_{c_k})\}$ of the clique complex $K$ of a graph $G$ (equation \ref{filtcrit}), each $p$-hole has a critical weight $w_{c{_{birth}}}$ which corresponds to its birth index and $w_{c{_{death}}}$ which corresponds to its death index, with $w_{c{_{birth}}} < w_{c{_{death}}}$.

Persistence diagram D$(f)$ for a $d$-dimensional simplicial complex $K$ is the collection of points in $\mathbb{R}^2$ whose first and second coordinates, $x$ and $y$, respectively, correspond to the birth weight and death weight of a $p$-hole where $0 \leq p \leq d$ \cite{Cohen-Steiner2007}. Since two different holes can have the same birth and death weights, each point in the persistence diagram has a corresponding multiplicity, we refer the reader to the SI Appendix for more details. Thus, the persistence diagram is a multiset of points in $\mathbb{R}^2$. The persistence of a $p$-hole which has birth and death weights, $w_{c{_{birth}}}$ and $w_{c{_{death}}}$, respectively, is defined as $w_{c{_{birth}}} - w_{c{_{death}}}$. Thus, the persistence diagram for a clique complex $K$ corresponding to a graph $G$ is a compact representation of the persistent homology of a network.

Given two persistence diagrams $X$ and $Y$ (which may correspond to two different networks), the $\infty$-Wasserstein distance between $X$ and $Y$, also
known as the \textit{bottleneck distance} \cite{Cohen-Steiner2007}, is defined as follows:
\begin{equation}
\label{botdist}
W_\infty(X,Y) =  \inf_{\eta:X \rightarrow Y} \text{sup}_{x \in X} || x - \eta(x) ||_{\infty}.
\end{equation}
Similarly, given two persistence diagrams $X$ and $Y$, the $q$-Wasserstein distance \cite{Kerber2017} between $X$ and $Y$ is defined as follows:
\begin{equation}
W_q(X,Y) = \bigg[ \inf_{\eta:X \rightarrow Y} \sum_{x \in X} || x - \eta(x) ||_{\infty}^q \bigg]^{\frac{1}{q}}.
\end{equation}
In the above equations, $\eta$ ranges over all bijective maps from $X$ to $Y$,
and given $(a,b) \in \mathbb{R}^2$, $||(a,b)||_\infty = \text{max} \{ |a|, |b|
\} $ is the $L_\infty$ norm. In this work, we use Dionysus 2  package (\url{http://www.mrzv.org/software/dionysus2/}) to compute the Wasserstein distance between two persistence diagrams corresponding to two different model networks (See results section). Note that it is not generally true that two persistence diagrams $X$ and $Y$ have the same number of off-diagonal points, i.e., features with non-zero persistence, and we refer the readers to Kerber \textit{et al.} \cite{Kerber2017} for details on circumventing this issue and further information regarding how the computation of the Wasserstein distance is reduced to a bipartite graph matching problem in Dionysus 2 package. We remark that the bottleneck distance between two persistence diagrams which are subsets of the unit square is in the range 0 to 1.

Stability of persistence diagrams with respect to small changes in the discrete Morse function is a key property of persistent homology. The first such stability theorem was given by Cohen-Steiner \textit{et al.} \cite{Cohen-Steiner2007}, and later refined by Chazal \textit{et al.} \cite{Chazal2008}. In the SI Appendix, using results from Chazal \textit{et al.}\cite{Chazal2008}, we give a stability result for persistence diagrams of discrete Morse functions with respect to the bottleneck distance.


\section*{Network datasets}

\noindent \textbf{Model networks}. We have investigated the following models of unweighted and undirected networks, namely, the Erd\"{o}s-R\'{e}nyi (ER)
\cite{Erdos1961}, the Watts-Strogatz (WS) \cite{Watts1998}, the Barab\'{a}si-Albert (BA) \cite{Barabasi1999} and the Hyperbolic Graph Generator
(HGG) \cite{Krioukov2010}. The ER model \cite{Erdos1961} is characterized by the property that the probability $p$ of the existence of each possible edge between any two vertices among the $n$ vertices in the graph $G$ is constant. The existence of edges in the ER model are independent of each other, and thus, the model produces random graphs $G(n,p)$ with average vertex degree $p(n-1)$. The WS model \cite{Watts1998} produces small-world graphs as follows. The WS model starts with an initial regular graph with $n$ vertices where each vertex is connected to its $k$ nearest neighbours. Next, the endpoint of each edge in the initial regular graph of the WS model is randomly chosen for rewiring based on a fixed rewiring probability $p$ and is rewired to another vertex in the graph which is chosen with uniform probability. The BA model \cite{Barabasi1999} produces scale-free graphs which are characterized by a degree distribution that follows a power law decay. The BA model utilizes a preferential attachment scheme to produce scale-free graphs. The BA model generates an initial graph of $m_0$ vertices, and then, at each successive iteration a new vertex is added with edges to $m$ already existing vertices which are chosen with probability proportional to their degree at that particular iteration. The iterations in the BA model cease when the graph has attained the requisite number $n$ of vertices. The HGG model \cite{Krioukov2010,Aldecoa2015} produces a random graph of $n$ vertices by initially fixing $n$ vertices to $n$ points on a hyperbolic disk. In the HGG model, the probability of existence of an edge between two vertices is proportional to the hyperbolic distance between the two points on the hyperbolic disk that correspond to these two vertices. By tuning the input parameter $\gamma$, the HGG model can produce either a hyperbolic or a spherical random graph \cite{Krioukov2010,Aldecoa2015}. Specifically, the HGG model produces hyperbolic random graphs for $\gamma \in [2, \infty)$ whereas spherical random graphs for $\gamma = \infty$.

\noindent \textbf{Real networks}. We have also studied seven real-world networks which are represented as unweighted and undirected graphs. We have considered two biological networks, namely, the \textit{Yeast protein interaction} network \cite{Jeong2001} with 1870 vertices and 2277 edges, and the \textit{Human protein interaction} network \cite{Rual2005} with 3133 vertices and 6726 edges. In both biological networks, each vertex represents a protein and an edge represents an interaction between the two proteins. We have considered two infrastructure networks, namely, the \textit{US Power Grid} network \cite{Leskovec2007} and the \textit{Euro road} network \cite{Subelj2011}. In the US Power Grid network, the 4941 vertices represent the generators, transformers and substations in the Western states of USA and the 6594 edges represent power links between them. The 1174 vertices of the Euro road network correspond to cities in Europe and the 1417 edges correspond to roads linking the cities. We have also studied the \textit{Email} network \cite{Guimera2003} of the University of Rovira i Virgili with 1133 vertices representing users and 5451 edges, each representing the existence of at least one Email communication between the two users corresponding to the vertices anchoring the edge. We have also studied the \textit{Route views} network \cite{Leskovec2007} which has 6474 autonomous systems as vertices and 13895 edges representing communication between the systems that are represented as vertices. We have considered a social network, the \textit{Hamsterster friendship} network \cite{Kunegis2013}, containing 1858 vertices which represent the users and 12534 edges which represent friendships between the users. Note that we omit self-loops while constructing the clique complex $K$ corresponding to the undirected graph $G$ of a real-world network.


\section*{Results and Discussion}

\subsection*{Algorithm to construct discrete Morse function on a simplicial
complex}

From an unweighted and undirected graph $G(\mathcal{V},\mathcal{E})$ with vertex set $\mathcal{V}$ and edge set $\mathcal{E}$, it is straightforward to construct a clique simplicial complex $K$ with dimension $d$ (See Theory section). Figure \ref{fig1} shows the construction of a clique complex starting from an example network. Given a simplicial complex $K$, its dimension $d$ and a non-negative real-valued function $g$ on the 0-simplices of $K$, the algorithm 1 assigns weights to any simplex in $K$, producing a discrete Morse function $f$ defined in equation \ref{discmor}. In the pseudocode of the algorithm 1, lines 2-6 initialize a variable \texttt{Flag}$[\alpha]$ for every simplex $\alpha$ in clique complex $K$ with the value 0. We remark that the variable \texttt{Flag}$[\alpha]$ associated with a simplex $\alpha$ in $K$ serves as a counter for the size of the set $U_{\alpha}$ defined in equation \ref{set-u}. Lines 7-9 assign weights to every $0$-simplex in $K$ based on the input non-negative function $g$. Lines 10-24 assign weights to $1$- or
higher-dimensional simplices in $K$ in a manner which is consistent with the definition in equation \ref{discmor} of a discrete Morse function. In summary, algorithm 1 outputs a discrete Morse function $f$ on $K$, and in SI Appendix, we present a rigorous proof for the following theorem which states
the same.

\vspace{5pt}
\noindent \textbf{Theorem.} \textit{Algorithm 1 produces a discrete Morse function $f$ on any simplicial complex $K$ of finite dimension $d$.}\\
\begin{algorithm}[H]
\caption{Algorithm to construct a discrete Morse function on a $d$-dimensional simplicial complex $K$}
\begin{algorithmic}[1]
\Function{DiscreteMorseFunction}{$K,d,g$}
\Statex
    \For{$p=0,\cdots,d$}                    \Comment{Initialize \texttt{Flag}
variable associated with each simplex in $K$}
	\For{each $p$-simplex $\alpha \in K$} 	
	   \State \texttt{Flag}$[\alpha]=0$		\Comment{The variable
\texttt{Flag} keeps track of the size of the set $U_{\alpha}$ for each simplex
$\alpha$}
	\EndFor
    \EndFor
    \Statex
    \For{each $0$-simplex $\alpha \in K$}  	\Comment{Assign weights to
$0$-simplices in $K$}
		\State $f(\alpha) = g(\alpha)$
	\EndFor
	\Statex
    \For{$p=1,\cdots,d$}
	\For{each $p$-simplex $ \alpha \in K$}  \Comment{Assign weights to
$p$-simplices in $K$ with $p \geq 1$}
		\State Let \texttt{Faces}[ ] be an array of all
$(p-1)$-dimensional faces of $\alpha$
		\State Sort \texttt{Faces}[ ] such that $f(\texttt{Faces}[i])
\geq f(\texttt{Faces}[i+1])$ for each $i \in \{0,1,\cdots,p-1\}$
		\State Let $\gamma_0 = \texttt{Faces}[0]$
		\State Let $\gamma_1 = \texttt{Faces}[1]$
		\If{ \texttt{Flag}$[\gamma_0] = 0$ and $f(\gamma_0) >
f(\gamma_1)$ }
			\State $f(\alpha) = (f(\gamma_0)+f(\gamma_1))/2$
			\State \texttt{Flag}$[\gamma_0] = 1$
		\Else
			\State $\epsilon =$ random$(0,0.5)$ \Comment{$\epsilon$
is  generated randomly during runtime using a uniform distribution on $(0,0.5)$}
			\State $f(\alpha) = f(\gamma_0) + \epsilon$
		\EndIf
	\EndFor
	\EndFor
	\Statex
	\State \Return $f$
\Statex
\EndFunction
\end{algorithmic}
\end{algorithm}

Given a simplicial complex $K$, its dimension $d$ and a discrete Morse function $f$ on $K$, the algorithm 2 in SI Appendix determines the weights of critical simplices in $K$. Given an unweighted and undirected graph $G$, we restrict the construction of clique complex $K$ by including simplices up to a maximum dimension $d$. Then, the algorithm 3 in SI Appendix creates the filtration of clique complex $K$ based on weights of critical simplices as described in the Theory section. In SI Table S1, we describe the role of key variables which appear in algorithms 1, 2 and 3. In SI Appendix, we also give a time complexity analysis for these algorithms.


\subsection*{Rationale for the choice of function on vertices}

In order to construct a discrete Morse function $f$ on clique complex $K$ corresponding to a graph $G$ using
our algorithm 1, a real-valued function $g$ has to be fixed on the $0$-simplices of $K$ (See lines 7-9 in
algorithm 1). Let $deg_{max}$ denote the maximum degree of a vertex in the graph $G(\mathcal{V},\mathcal{E})$.
Our choice for the function value on the vertices or $0$-simplices, $g: \mathcal{V} \rightarrow \mathbb{R}$,
is as follows:
\begin{equation}
\label{g-vertex}
g(v) = deg_{max} - \text{degree}(v) + \epsilon
\end{equation}
where degree$(v)$ is the degree of the vertex $v \in G$ and $\epsilon$ corresponding to each vertex is a random
number (noise) generated using the uniform distribution on the interval $(0,0.5)$.

In the Theory section, we had highlighted the Theorem 2.11 by Forman \cite{Forman2002} which gives a lower bound
on the number of critical $p$-simplices, $m_p$, in a simplicial complex $K$ as the $p$-Betti number $\beta_p$.
The choice of the real-valued function $g$ in algorithm 1 plays a key role in determining if $m_p$ is close to
the theoretical minimum $\beta_p$ stated above. In the Theory section, we have shown that the number of critical
simplices determines the effective number of filtration weights to study the persistent homology of a clique
complex (See equation \ref{filtcrit}). This motivated our choice for the real-valued function (equation
\ref{g-vertex}) which determines the weights of $0$-simplices, and the rationale for this choice is as follows.

Ignoring the noise term $\epsilon$ in equation \ref{g-vertex}, the reader can discern our intuition for choosing
the function $g(v) = deg_{max} - \text{degree}(v)$ for any vertex $v$ in $G$ with the following example. Consider
the simple example of the clique complex $K$ corresponding to a graph $G$ in figure \ref{fig1}. Here, we would
like to obtain a discrete Morse function $f$ on $K$ such that the number of critical simplices is close to the
theoretical minimum. This requirement applies to simplices of any dimension in $K$, and in the context of this
example, we would like the number of critical $1$-simplices (edges) to be as close as possible to the $1$-Betti
number $\beta_1$ of $K$. Note that $\beta_1=1$ for the example clique complex $K$ in figure \ref{fig1}.

Let us now examine lines 11-23 in algorithm 1. Consider any edge $e_{vw}=[v,w]$ such that $g(v) > g(w)$. While
assigning the function value to the edge $e_{vw}$ in algorithm 1, the edge $e_{vw}$ and the vertex $v$ are
guaranteed to be not critical provided that the \textit{if condition} in the line 16 is satisfied. This is a
consequence of the definition of a critical simplex (See equation \ref{critsimp}). Thus, we would like to force
this \textit{if condition} to be True for as many edges as possible. Moreover, once the function value of the
$1$-simplex $e_{vw}$ is set, we set the variable \texttt{Flag}$[v]$ to 1 in line 18, and this subsequently
forces the \textit{if condition} in the line 16 to fail for all other edges $e_{vz} = [v,z]$ in the graph that
contain $v$ and have function value $g(v)>g(z)$.

Let us now examine the edge $e_{78}=[v_7, v_8]$ in figure \ref{fig1} which is anchored by vertices $v_7$ and $v_8$
with degree 5 and 1, respectively. As the degree of a vertex gives the number of edges that contain the vertex,
$v_7$ is part of 4 other edges apart from $e_{78}$ while $v_8$ is part of only the edge $e_{78}$. Suppose $e_{78}$
is the first edge chosen for the function assignment in line 11 of algorithm 1 and both \texttt{Flag[$v_7$]} and
\texttt{Flag[$v_8$]} for the anchoring vertices are $0$. We would then prefer that the \textit{if condition} in
the line 16 is satisfied for $e_{78}$, and as $v_7$ is part of 4 other edges apart from $e_{78}$ while $v_8$ is
part of only $e_{78}$, ideally \texttt{Flag[$v_8$]} is set to 1 instead of \texttt{Flag[$v_7$]}, in other words,
we need the function value $g(v_8) > g(v_7)$. We emphasize that this choice of $v_8$ over $v_7$ prevents the forced
failure (described in the previous paragraph) of the \textit{if condition} for the 4 other edges apart from $e_{78}$
that contain $v_7$.

The above example suggests a need for a function $g$ on the vertices that has an inverse relationship with the degree
of the vertices. Hence, our choice $g(v) = deg_{max} - \text{degree}(v)$ provides a simple and effective solution for
the above requirement. As a consequence of this choice for the function $g$, the weights assigned to the simplices by
algorithm 1 reflect the degree of its constituent vertices. The degree of a vertex can be thought of as a measure of
its importance in the network. Hence, the intuition behind the assignment of weights to simplices by our method is to
have an inverse or opposite relationship between `weight' and `importance' of a simplex while simultaneously satisfying
the definition of a discrete Morse function. This inverse relationship instead of a proportional
relationship between `importance' and `weight' plays a key role in our filtration scheme which is based on the sequence
of level subcomplexes corresponding to the increasing sequence of critical weights (See Theory section and algorithm 3
in SI Appendix). Thus, within the constraint of ensuring that the definition for a simplicial complex is satisfied at
each stage of the filtration, our scheme prioritizes the addition of simplices with higher `importance' at earlier
stages of the filtration due to their lower weights.

We now provide a rationale for the addition of a random noise $\epsilon$ in equation \ref{g-vertex}. As reasoned above,
we would like to force the \textit{if condition} in line 16 of algorithm 1 to be True for as many edges as possible.
Consider an edge $e_{vw}=[v,w]$ such that $degree(v) = degree(w)$. The absence of a random noise $\epsilon$ in equation
\ref{g-vertex} forces $g(v)=g(w)$. Thus, irrespective of the state of \texttt{Flag}$[v]$ and \texttt{Flag}$[w]$, the
\textit{if condition} fails. This implies that the set $V_{e_{vw}}$ (See equation \ref{set-v}) corresponding to the
edge $e_{vw}$ is forced to be empty since the only other possibility is $|V_{e_{vw}}| = 2$ which cannot be the case
because algorithm 1 produces a discrete Morse function (See Theorem and equation \ref{discmor}). Thus, provided $e_{vw}$
is not a face of any higher dimensional simplex, $e_{vw}$ would be a critical simplex irrespective of the state of
\texttt{Flag}$[v]$ and \texttt{Flag}$[w]$. Hence, we would like $g(v) \neq g(w)$ while also retaining the inverse
relationship of the function with the degree. Thus generating a small random noise $\epsilon$ in the range $(0,0.5)$
for each vertex as in equation \ref{g-vertex} provides a simple resolution. We remark that the above argument can be
generalized to higher-dimensional simplices, and thus, provides the intuition for the addition of noise $\epsilon$ in
line 21 of algorithm 1.

We remind the readers that our initial motivation was not to develop a scheme to construct the optimal discrete Morse
function on a clique complex corresponding to a graph. Rather, our main goal is to develop a systematic filtration scheme
to study persistent homology in unweighted and undirected networks. In fact, constructing an optimal discrete Morse
function in the general case has been shown to be MAX-SNP Hard \cite{Lewiner2003}. The primary utility of our scheme
is to create a filtration by assigning weights to simplices in the clique complex $K$ of a graph $G$. However, we next
report our empirical results from an exploration of model and real-world networks which underscore the following.
Although our scheme is not optimal in the sense of minimizing the number of critical simplices, in practice, it achieves
near-optimal results in several model and real-world networks (Table \ref{critical-counts}). Hence, our scheme based on
discrete Morse theory reduces the number of filtration steps and increases the applicability of persistent homology to
study complex networks.


\begin{table}[ht]
\caption{The table lists the number of $p$-simplices ($n_p$), the number of critical $p$-simplices ($m_p$) and the $p$-Betti number $\beta_p$ for clique
complexes corresponding to model and real networks. Note that the dimension $p$ of simplices ranges from 0 to 3. In case of model networks, the statistics is reported for a particular instance of ER graph with $n=1000$ and $p=0.004$, WS graph with $n=1000$, $k=4$ and $p=0.5$, BA graph with $n=1000$ and $m=2$, Spherical random graph produced from HGG model with $n=1000$, $T=0$, $k=4$ and $\gamma=\infty$, and Hyperbolic random graph produced from HGG model with $n=1000$, $T=0$, $k=4$ and $\gamma=2$. For the statistics corresponding to a more comprehensive list of model networks with different parameters, we refer the readers to SI Table S2. Note that we omit self-loops in the real networks considered here.}
\label{critical-counts}	
\begin{adjustbox}{width=1\textwidth}
\begin{tabular}{|l|c|c|c|c|c|c|c|c|c|c|c|c|}
\hline
\textbf{Network} & $n_0$ & $m_0$ & $\beta_0$ & $n_1$ & $m_1$ & $\beta_1$ & $n_2$
& $m_2$ & $\beta_2$ & $n_3$ & $m_3$ & $\beta_3$ \\
\hline
ER model with $n=1000$ and $p=0.004$  & 1000 & 90 & 21    & 2007 & 1090 & 1021
& 7 & 0 & 0 & 0 & 0 & 0 \\
WS model with $n=1000$, $k=4$ and $p=0.5$  & 1000 & 123 & 1    & 2000 & 991 &
864  & 137 & 5 & 0 & 0 & 0 & 0 \\
BA model with $n=1000$ and $m=2$  & 1000 & 8 & 1      & 1996 & 949 & 942  & 55 &
0 & 0  & 0 & 0 & 0 \\
Spherical model with $n=1000$, $T=0$, $k=4$  and $\gamma=\infty$   & 1000 & 172
& 126  & 2028 & 118 & 0 & 2029 & 180 & 0 & 1321 & 554 & 446 \\
Hyperbolic model with $n=1000$, $T=0$, $k=4$ and $\gamma=2$  & 1000 & 144 & 144
& 2593 & 20 & 0 & 5440 & 426 & 0 & 11456 & 8159 & 7753 \\
\hline
US Power Grid	& 4941	& 573	& 1 & 6594	& 1671	& 1080 & 651 & 21 &	
0 &	90 & 15 &	13 \\
Email communication	& 1133	& 6	& 1 &	5451 & 1694	& 1186 &	
5343 & 	871	& 53 &	3419 &	1577 &	1262\\
Route views 	& 6474	& 17	& 1 & 12572	& 2459	& 2157 & 6584  & 627 &	
19 & 5636 &	3335 &	3013\\
Yeast protein interaction 	& 1870	& 272	& 173 & 2203	& 424	& 318 &
222	& 12 &	0  & 41 & 12 & 7\\
Hamsterster frienship	& 1858	&33	& 23 & 12534	& 4484	& 2970 & 16750 &	
5324 & 1880 &	10015 &	4814 &	2874\\
Euro road	& 1174	& 213	& 26 & 1417	& 425	& 237	& 32	 & 1	
& 0 &	0 &	0 &	0 \\
Human protein interaction	& 3133	& 269	& 210 & 6149	& 2454	& 2298 &
1047	& 109 &	1 &	142	& 35 & 24\\
\hline
\end{tabular}
\end{adjustbox}
\end{table}
\begin{table}[ht]
\centering
\caption{The table lists the value of optimality indicator $\mu$ for various model and real networks analyzed here. Equation \ref{mu-def} gives the
definition of $\mu$ which is an indicator for the proximity of our algorithm to the optimal case. The value of $\mu$ ranges from $0$ to $1$ with $\mu=1$
indicating that our algorithm achieves exactly the theoretical minimum number of critical simplices while $\mu=0$ indicating that all simplices are critical. For model networks, the value reported in this table is the average of $\mu$ across 10 samples of each model network for a chosen set of parameter values along with the corresponding standard deviations.}
\label{opt-indicator}	
\begin{tabular}{|l|c|}
\hline
\textbf{Network} & \textbf{Optimalitiy indicator $\mu$}\\
\hline
ER model with $n=1000, p=0.004$	& 0.924 $\pm$ 0.004 \\
ER model with $n=1000, p=0.006$	& 0.947 $\pm$ 0.004 \\
ER model with $n=1000, p=0.008$	& 0.959 $\pm$ 0.006 \\
WS model with $n=1000, k=4, p=0.5$ &	0.890 $\pm$ 0.003 \\
WS model with $n=1000, k=6, p=0.5$ &	0.917 $\pm$ 0.007 \\
WS model with $n=1000, k=8, p=0.5$ &	0.906 $\pm$ 0.003 \\
BA model with $n=1000, m=2$	& 0.989 $\pm$ 0.003 \\
BA model with $n=1000, m=3$	& 0.985 $\pm$ 0.004 \\
BA model with $n=1000, m=4$	& 0.964 $\pm$ 0.006 \\
Spherical model with $n=1000, \gamma=\infty, T=0, k=4$	& 0.925 $\pm$ 0.007 \\
Spherical model with $n=1000, \gamma=\infty, T=0, k=6$	& 0.901 $\pm$ 0.003 \\
Spherical model with $n=1000, \gamma=\infty, T=0, k=8$	& 0.887 $\pm$ 0.003 \\
Hyperbolic model with $n=1000, \gamma=2, T=0, k=4$	& 0.939 $\pm$ 0.013 \\
Hyperbolic model with $n=1000, \gamma=2, T=0, k=6$	& 0.927 $\pm$ 0.007 \\
Hyperbolic model with $n=1000, \gamma=2, T=0, k=8$	& 0.921 $\pm$ 0.005 \\
\hline
US Power Grid	& 0.893937 \\
Email communication	& 0.871847 \\
Route views	& 0.952140 \\
Yeast protein interactions	& 0.942157 \\
Hamsterster friendship	& 0.793236 \\
Euro road	& 0.840678 \\
Human protein interaction & 0.957924 \\
\hline
\end{tabular}
\end{table}

\begin{figure*}
\includegraphics[width=.7\columnwidth]{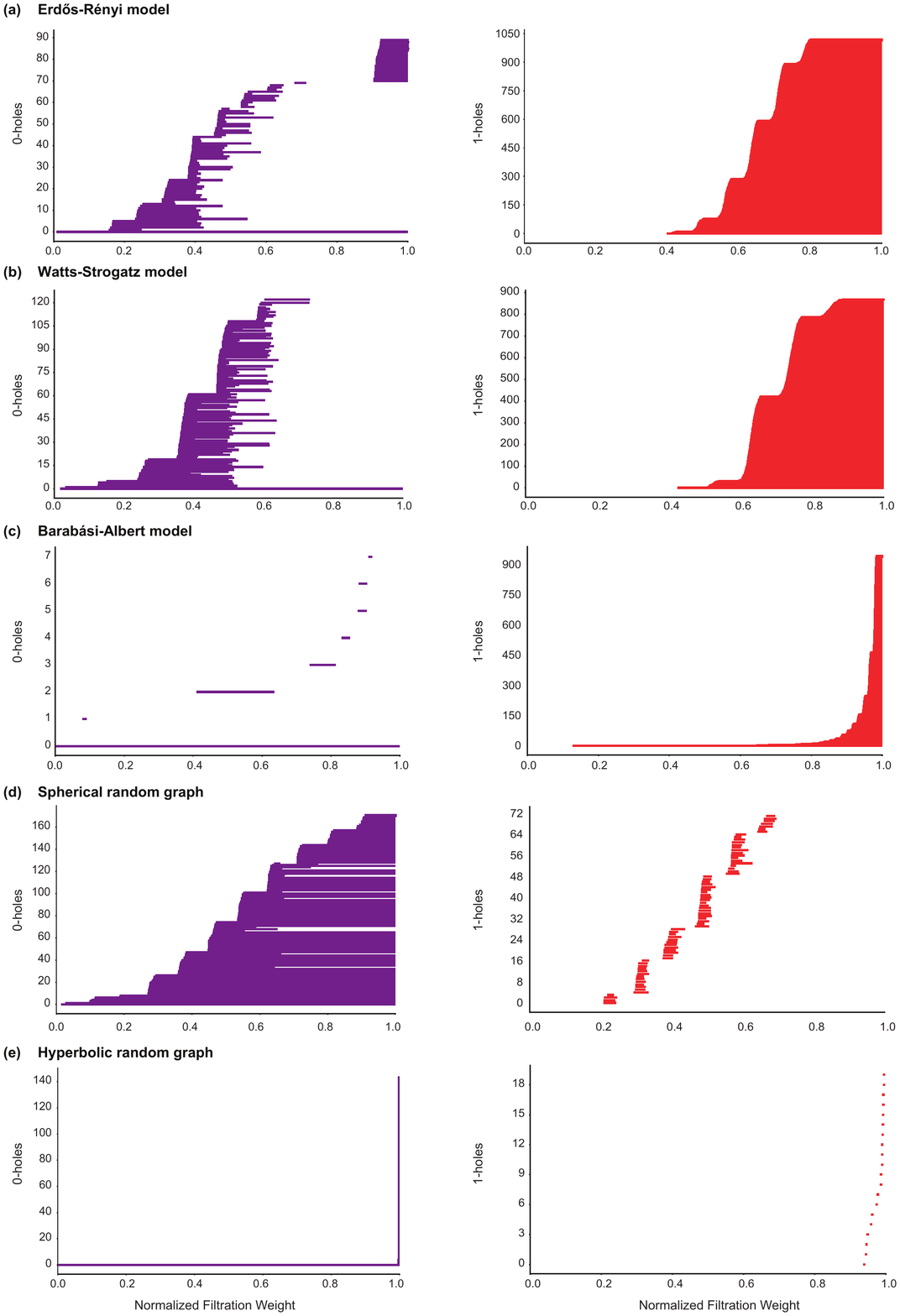}
\caption{Barcode diagrams for $H_0$ and $H_1$ in model networks. (a) ER model with $n=1000$ and $p=0.004$. (b) WS model with $n=1000$, $k=4$ and $p=0.5$. (c) BA model with $n=1000$ and $m=2$. (d) Spherical random graphs produced from HGG model with $n=1000$, $T=0$, $k=4$ and $\gamma=\infty$. (e) Hyperbolic random graphs produced from HGG model with $n=1000$, $T=0$, $k=4$ and $\gamma=2$.}
\label{bar-model}
\end{figure*}

\begin{figure*}
\includegraphics[width=.7\columnwidth]{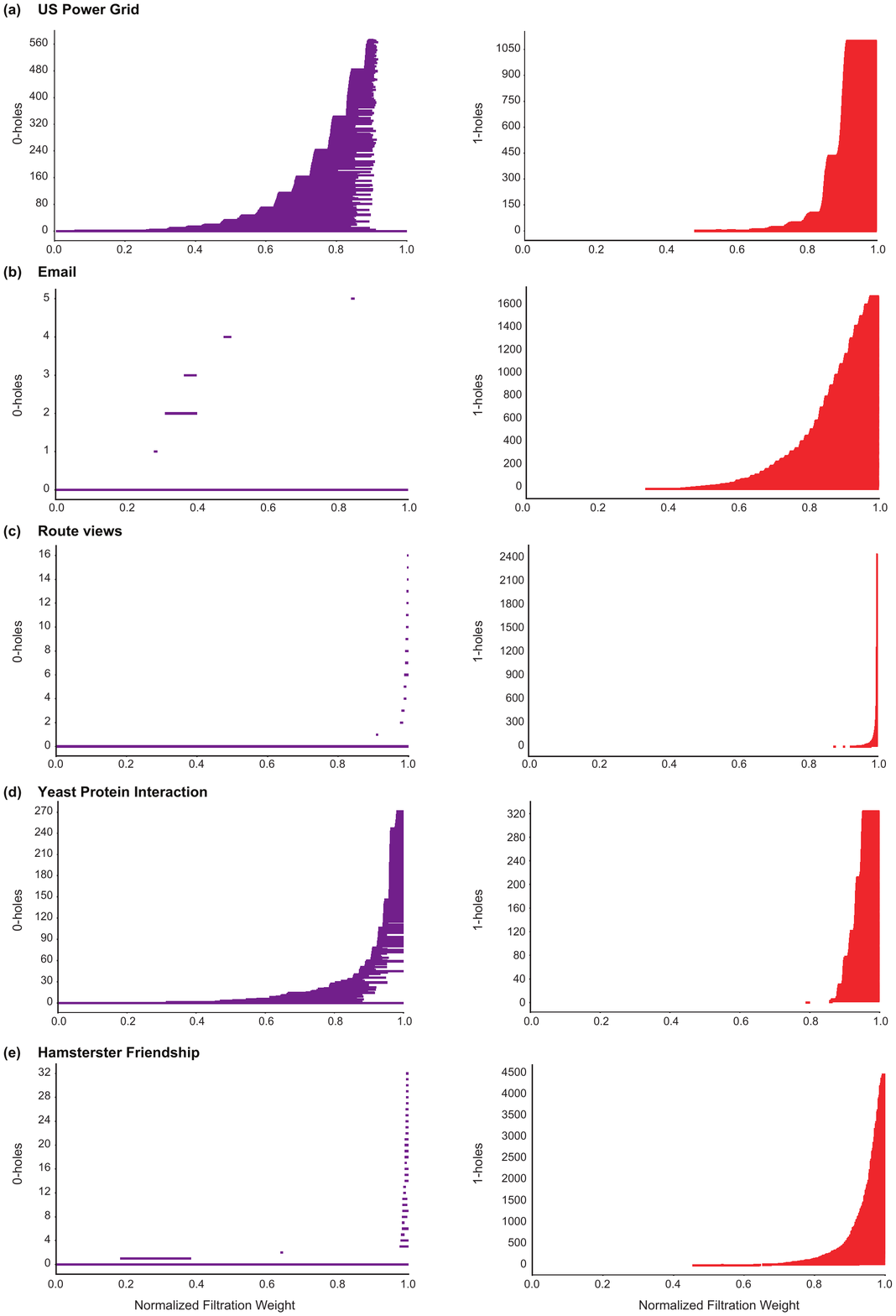}
\caption{Barcode diagrams for $H_0$ and $H_1$ in real networks. (a) US Power Grid. (b) Email communication. (c) Route views. (d) Yeast protein interaction. (e) Hamsterster friendship. }
\label{bar-real}
\end{figure*}

\begin{figure*}
\includegraphics[width=.7\columnwidth]{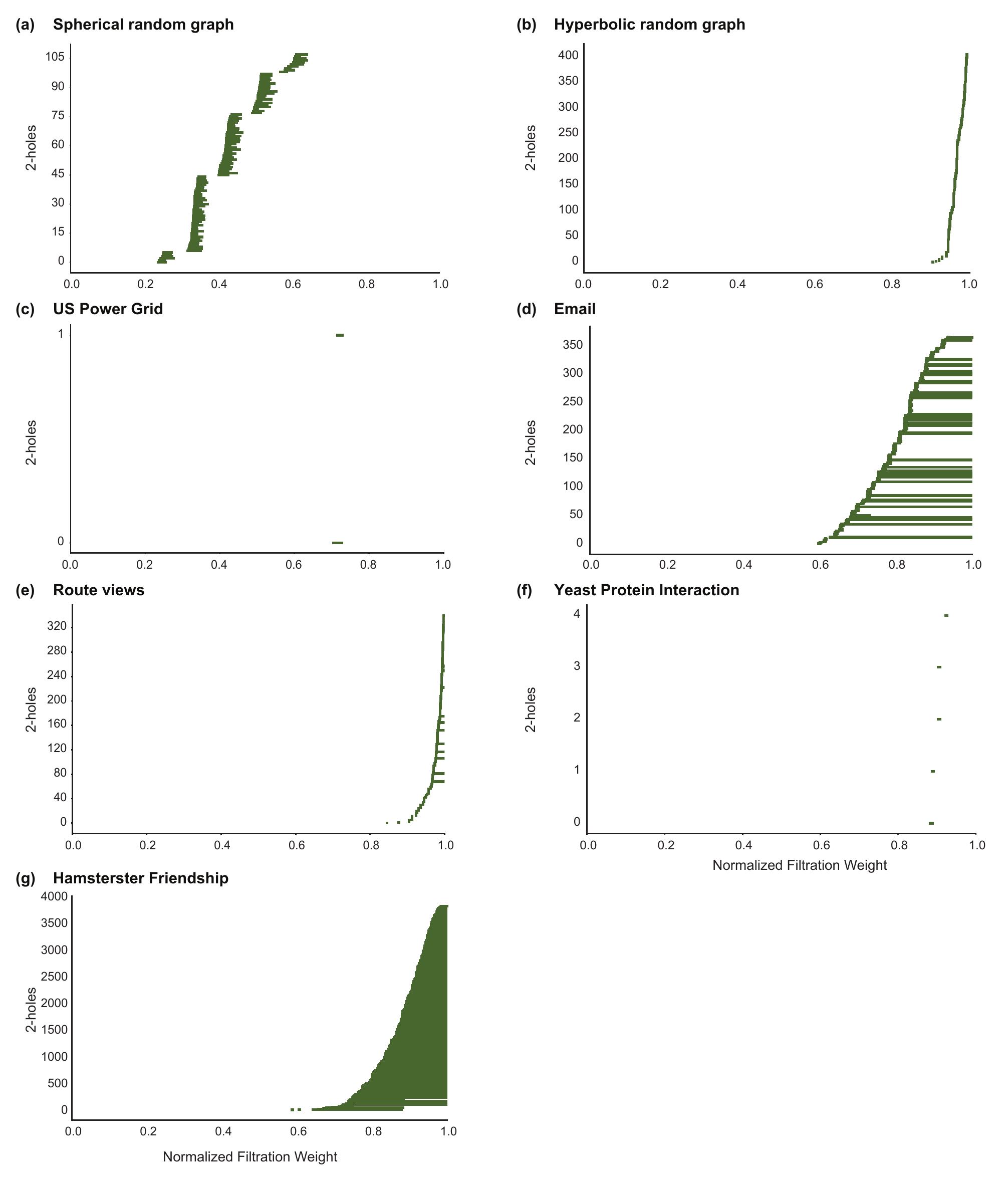}
\caption{Barcode diagrams for $H_2$ in model and real networks. (a) Spherical random graphs produced from HGG model with $n=1000$, $T=0$, $k=4$ and $\gamma=\infty$. (b) Hyperbolic random graphs produced from HGG model with $n=1000$, $T=0$, $k=4$  and $\gamma=2$. (c) US Power Grid. (d) Email
communication. (e) Route views. (f) Yeast protein interaction. (g) Hamsterster friendship.}
\label{h2}
\end{figure*}

\begin{figure*}
\includegraphics[width=.7\columnwidth]{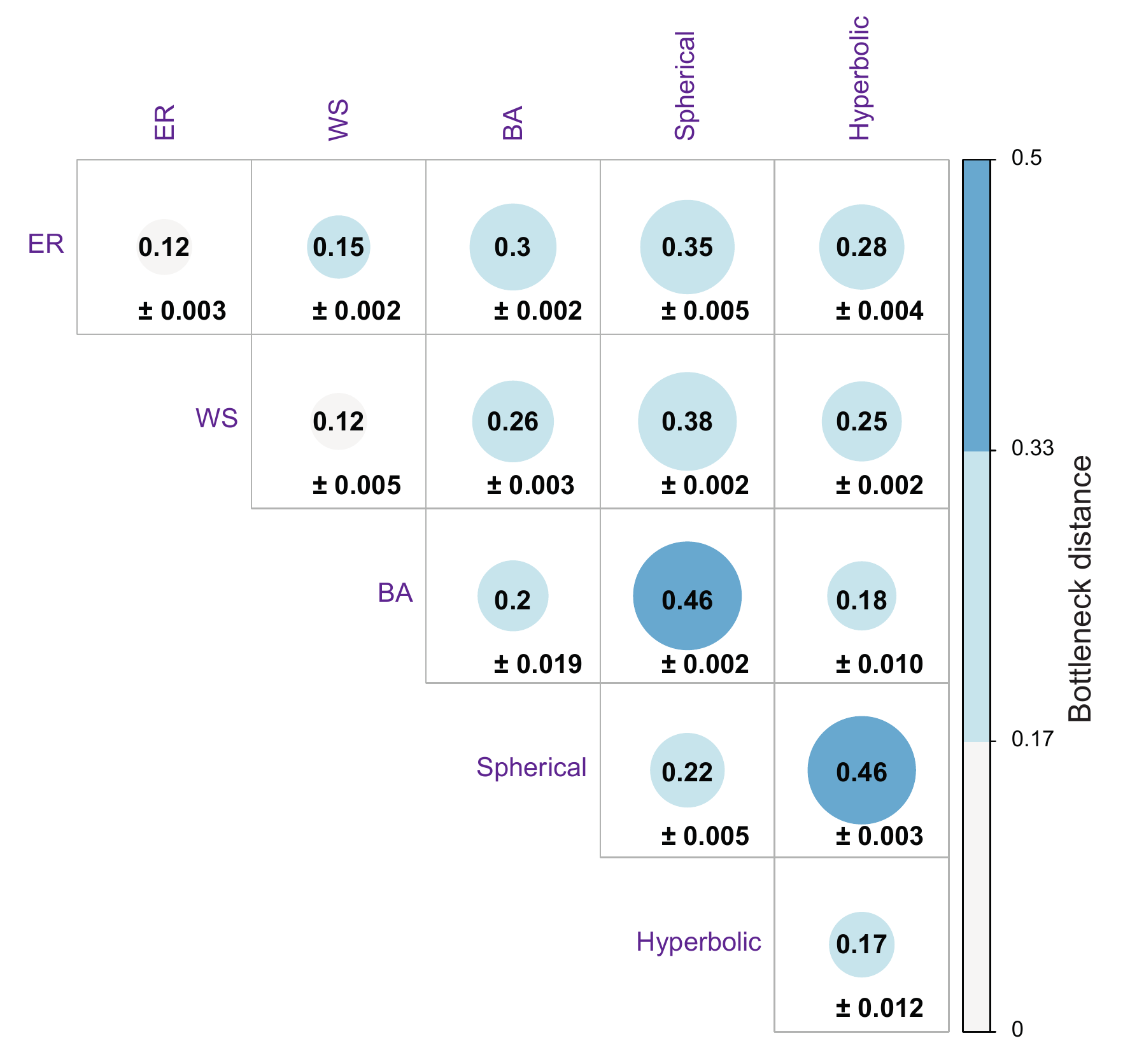}
\caption{Bottleneck distance between persistence diagrams of model networks, namely, ER model with $n=1000$ and $p=0.004$,  WS model with $n=1000$, $k=4$ and $p=0.5$, BA model with $n=1000$ and $m=2$, Spherical random graphs produced from HGG model with $n=1000$, $T=0$, $k=4$ and $\gamma=\infty$, and Hyperbolic random graphs produced from HGG model with $n=1000$, $T=0$, $k=4$ and $\gamma=2$. For each of the five model networks, 10 random samples are generated by fixing the number of vertices $n$ and other parameters of the model. We report the distance (rounded to two decimal places) between two different models as the average of the distance between each of the possible pairs of the 10 sample networks corresponding to the two models along with the standard error.}
\label{bottleneck}
\end{figure*}

\subsection*{Application to model and real networks}

Given a model or real network $G$, we limit our study of persistent homology to the $3$-dimensional clique simplicial complex $K$ corresponding to $G$. In other words, during the construction of the clique complex, we only include $p$-simplices which have dimension $0 \leq p \leq 3$ (See Theory section). On this $3$-dimensional clique complex $K$, we create the corresponding filtration based on the assigned weights to simplices using discrete Morse theory. In SI Appendix, we present algorithm 3 which outlines the procedure to compute the filtration weights of simplices. Thereafter, we make use of GUDHI \cite{Maria2014}, a C++ based library for Topological Data Analysis (\url{http://gudhi.gforge.inria.fr/}) to study the persistent homology of this filtration of $K$.

Each hole of any dimension in $K$ has a corresponding \textit{birth} filtration weight and a \textit{death} filtration weight (See Theory section). We normalize the \textit{birth} and \textit{death} filtration weights of all holes in $K$ by dividing with $w_N = 1 + \text{max} \{ f(\alpha) \: | \: \alpha \in K \}$. In other words, $w_N$ is $1$ plus the maximum value among weights assigned to the simplices in $K$. We also make the convention that a normalized \textit{death} filtration weight of $1$ for a hole in $K$ represents that the particular hole never \textit{dies}.

A $H_p$-barcode diagram corresponding to a filtration of the clique complex $K$ is a graphical representation containing horizontal line segments, each of which represents a $p$-hole in $K$, plotted against the $x$-axis ranging from $0$ to $1$ which corresponds to the normalized filtration weights of simplices in $K$ \cite{Ghrist2008}. A horizontal line in the $H_p$-barcode diagram of $K$ is referred to as a \textit{barcode}. Thus, a barcode in the $H_p$-barcode diagram of $K$ which begins at a $x$-axis value of $w_1$ and ends at a $x$-axis value of $w_2$ represents a $p$-hole in $K$ whose \textit{birth} and \textit{death} weights are $w_1$ and $w_2$, respectively. In figures \ref{bar-model}, \ref{bar-real} and \ref{h2}, we display the barcode diagrams for model and real-world networks analyzed here.

\subsubsection*{Model Networks}

In this work, we have investigated the persistent homology of unweighted and undirected graphs corresponding to five model networks, namely, ER, WS, BA, hyperbolic random graphs and spherical random graphs. We have considered model networks with 1000 vertices and expected average degree 4, 6 and 8. In main text, we report results for model networks with expected average degree 4, and in SI, those with expected average degree 6 and 8.

The $H_0$ barcode diagram of BA networks indicate a low number of $0$-holes in BA networks across the entire filtration (Figure \ref{bar-model} and SI Figures S1-S2). A standard result in algebraic topology \cite{Munkres2018} gives that the $0$-Betti number of a simplicial complex $K$ is equal to the number of connected components in $K$. In other words, the above observation indicates that the scale-free BA network has a strong tendency to maintain a low number of connected components during filtration (Figure \ref{bar-model} and SI Figures S1-S2). In contrast, both the random ER network and
small-world WS network have a relatively high number of connected components at initial phase of the filtration and then progress towards a more connected network at later stages of the filtration (Figure \ref{bar-model} and SI Figures S1-S2). This indicates that the simplices in the clique complex which are key to the connectivity of the model networks are introduced very early into the filtration for the scale-free BA network while this is not the case for the random ER network or small-world WS network. The $H_1$ barcode diagram of BA networks also indicate late introduction of $1$-holes during filtration in contrast to both ER and WS networks where $1$-holes appear across a wider range of the filtration (Figure \ref{bar-model} and SI Figures S1-S2). Moreover, in ER and WS networks, it is interesting to observe that the $1$-holes start to appear at roughly the same stage of filtration which corresponds to a sharp reduction in the number of connected components (Figure \ref{bar-model} and SI Figures S1-S2).

In contrast to ER, WS and BA networks, the spherical and hyperbolic networks are characterized by a relatively high $0$-Betti number $\beta_0$ and a low $1$-Betti number $\beta_1$ (Table \ref{critical-counts} and SI Table S2). Simply stated, this observation on the magnitude of $\beta_0$ indicates that both hyperbolic and spherical networks have a higher number of connected components in comparison to the ER, WS and BA networks of similar size, i.e., number of vertices, and average vertex degree (Table \ref{critical-counts} and SI Table S2). Although, both spherical and hyperbolic networks exhibit a higher number of connected components, they differ from each other with respect to the evolution of these connected components during filtration (Figure \ref{bar-model} and SI Figures S1-S2). The hyperbolic model maintains a relatively low number of connected components until very late in the filtration wherein there is a sharp increase in the number of connected components (Figure \ref{bar-model} and SI Figures S1-S2). This is in contrast with the behavior of the $H_0$ barcode diagram of the spherical model which exhibits a more distributed evolution of connected components during filtration (Figure \ref{bar-model} and SI Figures S1-S2). In addition, the low $\beta_1$ for spherical and hyperbolic networks conveys the lack of $1$-holes in these networks. A possible reason for this observation is the incidence of higher number of $2$-simplices in the clique complex $K$ of the spherical and hyperbolic networks in comparison to the ER, WS and BA networks (Table \ref{critical-counts} and SI Table S2). Note that the formation of a $2$-simplex can potentially fill in a $1$-hole, and thus, result in a low value for $\beta_1$ (See Theory section). Such a behaviour is also seen in the $H_2$ barcode diagrams of spherical and hyperbolic networks wherein the $2$-holes have very short persistence since the addition of $3$-simplices successively fill in
the $2$-holes (Figure \ref{h2} and SI Figure S3). The $H_3$ barcode diagrams of spherical and hyperbolic networks (See SI Figures S3-S4) also indicate a clear difference in the evolution of their corresponding topological features during the filtration. The $H_2$ and $H_3$ barcode diagrams of ER, WS and BA networks do not provide any insight into network structure primarily due to a lack of higher-order correlations in these model networks that is essential for the formation of $2$-holes and $3$-holes.

A visual inspection of the barcode diagrams for the five model networks (Figures \ref{bar-model}, \ref{h2} and SI Figures S1-S4) suggests that the different models can be distinguished based on their persistent homology. In Theory section, we had introduced the bottleneck distance which can be employed to quantify the differences between persistence diagrams from filtration of clique complexes corresponding to different model networks. Recall that the persistence diagram of a $d$-dimensional simplicial complex $K$ is a compact representation of the persistent homology of $K$ which encompasses topological information across all $d$ dimensions. Figure \ref{bottleneck} and SI Table S3 give the bottleneck distance between different model networks with the same number of vertices and similar average vertex degree. For each of the five model networks, 10 random samples are generated by fixing
the number of vertices $n$ and other parameters of the model. In figure \ref{bottleneck} and SI Table S3, we report the distance between two different models as the average of the distance between each of the possible pairs of the 10 sample networks corresponding to the two models along with the standard error. We find a relatively higher distance between a random instance of a BA network and a random instance of a ER network with the same number of vertices and similar average degree (Figure \ref{bottleneck} and SI Table S3). Similarly, we observe a relatively higher distance between a random instance of a BA network and a random instance of a WS network with similar size and average degree. In contrast, a relatively lower average distance is observed between a random instance of a ER network and a random instance of a WS network with similar size and average degree (Figure \ref{bottleneck} and SI Table S3). These observations indicate a similarity between networks generated by ER and WS models in terms of their corresponding persistence diagrams and also show that the BA model exhibits topological properties that are different from the other two model networks with the same number of vertices and similar average degree. Moreover, the average distance between a random instance of a spherical network and a random instance of a hyperbolic network
with similar size and average degree is very high (Figure \ref{bottleneck} and SI Table S3). The last observation is a reflection of the differences in the persistent homology of the clique complexes corresponding to spherical and hyperbolic networks. Finally, the nature of differences observed between persistence diagrams of different model networks using bottleneck distance as shown in figure \ref{bottleneck} remain consistent if the 1-Wasserstein or 2-Wasserstein distance is employed in place of the bottleneck distance (data not shown).

Table \ref{critical-counts} and SI Table S2 list the empirical data on the number of $p$-simplices $n_p$, the number of critical $p$-simplices $m_p$, that our algorithm achieves and the $p$-Betti number $\beta_p$ of the clique complexes corresponding to the five model networks. In Table \ref{opt-indicator} we report a value $\mu$ which indicates the optimality of our algorithm with respect to reducing the number of critical simplicies for each model network that has been analyzed. The definition for $\mu$ is as follows.
\begin{equation}
\label{mu-def}
\mu = \frac{\sum\limits_{p=0}^{d}n_p -
\sum\limits_{p=0}^{d}m_p}{\sum\limits_{p=0}^{d}n_p -
\sum\limits_{p=0}^{d}\beta_p}
\end{equation}
Here $d$ is the dimension of the corresponding clique complex. The value $\mu$ corresponding to a particular discrete Morse function on the clique complex of a given network is an indicator of its optimality with respect to minimizing the number of critical simplices. The value of $\mu$ ranges from $0$ to $1$. $\mu=1$ indicates that the discrete Morse function achieves exactly the minimum number of critical simplices thereby corresponding to the most optimal situation, while $\mu=0$ indicates that all simplices are critical thereby corresponding to the least optimal case. Note that the value of $\mu$ for a particular network increases linearly with a decrease in the number of critical simplices. In Table \ref{opt-indicator}, for model networks, the value reported is the average of $\mu$ across 10 samples of each model network for a chosen set of parameter values along with the corresponding standard deviations. Based on the data presented in Table \ref{opt-indicator}, we report that our algorithm achieves near-optimal results in terms of reducing
the number of critical simplices for each of the five model networks analyzed. Moreover for these model networks, data presented in SI Table S2 underlines the close proximity of $m_p$ to the theoretical lower bound $\beta_p$ across all dimensions $p=0,1,2,3$ (See Theory section). We remark that the worst case scenario corresponds to all simplices in the clique complex being critical which is never attained by our algorithm in the analyzed networks.


\subsubsection*{Comparison of our method with dimension based filtration in model networks}

In SI Figures S5 and S6, we show the $H_0$, $H_1$, $H_2$ and $H_3$ barcode diagrams for model networks with expected average degree 4 obtained using the dimensional filtration scheme used in Horak \textit{et al.} \cite{Horak2009} We restrict our investigation to the three-dimensional clique complex while computing the barcode diagrams for model networks using the dimensional filtration scheme of Horak \textit{et al.} \cite{Horak2009} In SI Figures S5 and
S6, we normalize the filtration index to be in the range 0 to 1, and $p$-holes with normalized filtration index 1 indicate that they never die. Moreover, we also report in SI Table S4 and SI Figure S7, the bottleneck distances between persistence diagrams of model networks obtained by the dimensional filtration scheme used in Horak \textit{et al.} \cite{Horak2009}, and we find that the resultant barcode diagrams and the bottleneck distances between the persistence diagrams are inconclusive to distinguish between the five model networks which have similar size and average vertex degree.

The discrete Morse function produced by our algorithm is empirically near-optimal for the model networks analyzed here, in the sense of minimizing the number of critical simplices (See Table \ref{opt-indicator}). On the other hand, it should be noted that the dimension function used in Horak \textit{et al.} \cite{Horak2009} is the most non-optimal in this regard. Moreover, in terms of persistent homology, our methods based on the discrete Morse function constructed using algorithms 1, 2 and 3 have the distinct feature of being able to distinguish between various model networks with an efficient
filtration.

Although the dimension function used in Horak \textit{et al.} \cite{Horak2009} is algorithmically efficient in terms of having a low number of filtration steps, unlike our method it does not conclusively distinguish between different model networks based on the barcode diagrams and also the bottleneck distances between the corresponding persistence diagrams. Thus our discrete Morse function is a good candidate for applications in both, computational aspects of discrete Morse theory as well as in persistent homology of unweighted networks, and achieves a tradeoff between efficiency and
applicability.


\subsubsection*{Real networks}

In this work, we have investigated the persistent homology of unweighted and undirected graphs corresponding to seven real-world networks and the barcode diagrams for five of these real networks is shown in figures \ref{bar-real}, \ref{h2} and SI figure S4. Based on the $H_0$ barcode diagrams, the behavior of most real networks considered here can be broadly classified into two categories. Real networks such as the Email communication, the Hamsterster
friendship and the Route views exhibit a relatively low number of connected components across the entire range of filtration (Figure \ref{bar-real}). On the other hand, the two biological networks, namely, the Yeast protein interaction and the Human protein interaction, exhibit a sharp increase in the number of connected components at the later stages of filtration. The $H_0$ barcode diagrams for the US Power Grid and Euro road do not conform with
either of the above characterizations. The $H_0$ barcode diagram of the US Power Grid network reveals that though there exists only a single connected component at the end of the filtration, there are a considerable number of non-persisting connected components that appear and subsequently disappear during filtration (Figure \ref{bar-real}). The $H_0$ barcode diagram of the Euro road network shows a more distributed increase in the number of connected components (data not shown).

In the context of the $H_1$ barcode diagrams, the real networks considered here exhibit similar properties with $1$-holes appearing late in the filtration (Figure \ref{bar-real}). The $H_2$ barcode diagrams reveal a lack of $2$-holes with long persistence in both biological networks, as well as the Route views network, the Euro road network and the US Power Grid network (Figure \ref{h2}). In contrast, from the $H_2$ and $H_3$ barcode diagrams (Figure \ref{h2}, SI Figure S4) we find that the social network, Hamsterster friendship, and the Email communication network exhibit a relatively high number of $2$-holes and $3$-holes with longer persistence.

Table \ref{opt-indicator} lists the value of the optimality indicator $\mu$ (See Equation \ref{mu-def}) for each of the seven real networks analyzed here. This data indicates near-optimal performance of our algorithm with respect to minimizing the number of critical simplices for each of these seven real networks. Table \ref{critical-counts} also lists the empirical data on the number of critical $p$-simplices, $m_p$, that our algorithm achieves and the
$p$-Betti number $\beta_p$ of the clique complexes across each dimension $p$, corresponding to the seven real networks analyzed here.


\section*{Conclusions}

To conclude, we have proposed a systematic scheme based on discrete Morse theory to study the persistent homology of unweighted and undirected networks. Our methods leverage the concept of \textit{critical} simplices to permit a reduced filtration scheme while simultaneously admitting a finer inspection of the changes in topology across the filtration of a clique complex corresponding to an unweighted network. Moreover, our proposed algorithm to construct
a discrete Morse function on the clique complex of a simple graph achieves close to optimal number of critical simplices for several model and real networks that have been studied here. Furthermore, based on visual representations of persistent homology such as the barcode diagrams as well as quantitative information in the form of distance between persistence diagrams, our methods successfully distinguish various model networks that exhibit inherently different properties. This motivates the application of our methods to real-world networks. We report the results obtained for seven real-world networks that are well studied in the network science community and observe certain patterns in the evolution of their topological features across the filtration. For instance both biological networks, namely the Yeast protein interaction network and the Human protein interaction network exhibit similar characteristics with respect to the $H_0$, $H_1$, $H_2$ and $H_3$ barcode diagrams. Similarly, both the Email network and the Hamsterster friendship
network, exhibit shared features with respect to $H_0$, $H_1$, $H_2$ and $H_3$ barcode diagrams that vary from the characteristics of the two biological networks considered here. Our observations hint at the ability and possible applications of our methods to detect and classify real-world networks that are inherently different.

Future directions and ongoing work include examining the significance of critical simplices in the context of real-world networks. In other words, we aim to determine whether a critical edge in the context of discrete Morse theory holds any key significance when it is viewed as a link between two real entities in a real-world network. We also intend to explore the presence or absence of a correlation between the notion of critical simplices and network curvature. Since discrete Morse theory captures information about the Euler characteristic of the clique complex corresponding to a graph, the
presence of such a correlation could potentially signify a close relationship between the discrete curvature of a graph and its topology, much like in the case of smooth, compact surfaces wherein the Gauss-Bonnet theorem relates the Gaussian curvature of a surface to its Euler characteristic.


\subsection*{Acknowledgments}

We thank Amritanshu Prasad for fruitful discussions. A.S. would like to acknowledge support from the Max Planck Society, Germany, through the award of a
Max Planck Partner Group in Mathematical Biology, and I.R. from the Science and Engineering Research Board (SERB) of the Department of Science and Technology (DST) India through the award of a MATRICS grant [MTR/2017/000835]. We thank the anonymous reviewers for their comments which have helped improve this manuscript.

\subsection*{Author contributions}

H.K., I.R. and A.S. designed the study. H.K. performed the simulations. H.K., E.S., I.R. and A.S. analyzed results. H.K., I.R. and A.S. wrote the manuscript. All authors reviewed and approved the manuscript.





\begin{center}
\noindent{\Large \bf Supplementary Information (SI)\\ Appendix\\}
\end{center}
\renewcommand{\theequation}{A.\arabic{equation}}
\renewcommand{\thefigure}{A\arabic{figure}}
\setcounter{equation}{0}

\section*{Homology of a simplicial complex}

In this section, we provide an overview of the mathematical theory of homology groups of a simplicial complex.

We first describe a mathematical group which provides the machinery to represent paths in a simplicial complex. The $p$-chain group $C_p$ of a simplicial complex $K$ is the Abelian group generated by the oriented $p$-simplices in $K$ with coefficients in a field $\mathbb{F}$. Elements $c$ of the $p$-chain group $C_p$ are referred to as $p$-chains and have the form $c = \sum_i n_i \alpha_i$ where we use the same notation to represent both the oriented $p$-simplex and its corresponding generator in $C_p$ and $n_i$ are scalars from the field $\mathbb{F}$. Note that the identity element $0$ in $C_p$ is the unique $p$-chain for which all the coefficients $n_i$ are zero in $\mathbb{F}$. Note also that if two $p$-simplices, $\alpha$ and $\beta$, have the same vertex set but opposite orientations, then the generators of $C_p$ corresponding to the two simplices are inverse of each other (i.e., $\alpha=-\beta$). In figure 1 of the main text, the cycle of length 4 formed by the edges or 1-simplices, $[v_1,v_2]$, $[v_2,v_5]$, $[v_5,v_7]$ and $[v_1,v_7]$, can be represented as an element of the $1$-chain group which is $c=[v_1,v_2]+[v_2,v_5]+[v_5,v_7]-[v_1,v_7]$.

The boundary operator $\partial_p$ on the generator corresponding to an oriented $p$-simplex $\alpha = [v_0, v_1,\cdots,v_p]$ is defined as follows
\cite{Munkres2018}: 
\begin{equation}
\partial_p (\alpha) = \sum_{i=0}^p (-1)^{i}
[v_0,v_1,\cdots,\hat{v_i},\cdots,v_p],
\end{equation}
where $\hat{v_i}$ refers to the absence of $v_i$ in the $(p-1)$-simplex. By linear extension, the boundary operator $\partial_p$ on any element $c = \sum_i n_i \alpha_i$ of $p$-chain group $C_p$ gives:
\begin{equation}
\partial_p(c) = \sum_i n_i \partial_p(\alpha_i)
\end{equation}
Note that the boundary operator in the above equation maps a $p$-chain to a $(p-1)$-chain. In figure 1 of the main text, the boundary operator applied to
the $1$-chain $c=[v_0,v_1]$ representing an edge gives $v_1 - v_0$ while the boundary operator applied to the $1$-chain $c=[v_1,v_2]+[v_2,v_5]+[v_5,v_7]-[v_1,v_7]$ representing a cycle of length 4 gives 0.

This motivates the definition of $p$-cycles $Z_p$ and $p$-boundaries $B_p$. The $p$-cycles $Z_p$ are the elements of the $p$-chain group $C_p$ which are mapped to 0 by the boundary operator $\partial_p$, and thus:
\begin{equation}
Z_p = \text{ker}(\partial_p) = \{ c \in C_p \: | \: \partial_p(c) = 0 \: \}.
\end{equation}
The $p$-boundaries $B_p$ is defined as follows:
\begin{equation}
B_p = \text{img}(\partial_{p+1})  = \{ c \in C_p \: | \: \exists \ b \in
C_{p+1}, \partial_{p+1} (b) =c \}.
\end{equation}
Thus, $p$-boundaries $B_p$ are the elements of the $p$-chain group $C_p$ which also happen to be the boundary of an element in the $(p+1)$-chain group
$C_{p+1}$. Note that both $Z_p$ and $B_p$ are subgroups of the $p$-chain group $C_p$. It can be shown that the composition $\partial_{p} \circ \partial_{p+1} = 0$. This implies that $B_p$ is a subgroup of $Z_p$ \cite{Munkres2018}. Simply stated, the $p$-boundary operator $\partial_{p}$ when applied on a $p$-boundary gives 0. Hence, it follows that every $p$-boundary is a $p$-cycle but not necessarily vice versa.

The $p$-homology group is defined as \cite{Munkres2018}:
\begin{equation}
H_p = Z_p/B_p,
\end{equation}
where $Z_p/B_p$ is the quotient group \cite{Dummit2003} of $Z_p$ over $B_p$. We informally refer to the elements of $p$-homology group $H_p$ as $p$-holes. To provide an intuition for the definition of homology groups, a natural way to describe a $p$-hole would be to characterize it as a $p$-cycle which is not a $p$-boundary. In figure 1 of the main text, the $1$-cycle $c=[v_1,v_2]+[v_2,v_5]+[v_5,v_7]-[v_1,v_7]$ is a $1$-hole as it is not a $1$-boundary of any $2$-chain while the $1$-cycle $c=[v_2,v_3]+[v_3,v_4]+[v_4,v_2]$ is not a $1$-hole as it is the $1$-boundary of the $2$-simplex $[v_2,v_3,v_4]$. Thus, the concept of quotient groups provide the mathematical machinery to characterize such $p$-holes in simplicial complexes.

We remark that though it is customary to call $H_{p}$ a group, since we use coefficients from a field $\mathbb{F}$, it satisfies additional properties which make it a vector space over $\mathbb{F}$. The $p$-Betti number is defined as the dimension of the homology group $H_p$ viewed as a vector space over field $\mathbb{F}$ \cite{Munkres2018}. Informally, the $p$-Betti number $\beta_p$ represents the number of $p$-holes of the simplicial complex. We remark that the Euler characteristic of the clique complex $K$ with dimension $d$ corresponding to a graph $G$ is given by the alternating sum of Betti numbers
\cite{Munkres2018}, namely,
\begin{equation}
\chi(K) = \beta_0 - \beta_1 + \beta_2 - \cdots + (-1)^{d} \beta_d.
\end{equation}
In this work, we use the finite field $\mathbb{F}=\mathbb{Z}/2$, i.e., the field with two elements.


\section*{Proof of concept for Algorithm 1}

In this section, we present a rigorous proof for the following theorem.

\vspace{5pt}
\noindent \textbf{Theorem 1.} \textit{Algorithm 1 produces a discrete Morse function $f$
on any simplicial complex $K$ of finite dimension $d$.}\\

\noindent \textit{Proof.} Let $K$ denote a simplicial complex of dimension $d$. Recall that for the function $f$ on $K$ which is constructed by algorithm 1, for each simplex $\alpha^{p} \in K$, the two sets $U_{\alpha}$ and $V_{\alpha}$ are defined as follows:
\begin{eqnarray}
\nonumber U_{\alpha} = \{ \beta^{p+1} \: | \: \alpha^p < \beta^{p+1} \text{ and } f(\beta) \leq f(\alpha) \} \\
\nonumber V_{\alpha} = \{ \gamma^{p-1} \: | \: \gamma^{p-1} < \alpha^p \text{ and } f(\alpha) \leq f(\gamma) \}
\end{eqnarray}
\noindent To prove that $f$ is a discrete Morse function we need to show that for each simplex $\alpha \in K$, both $\vert V_\alpha \vert \leq 1$ and $\vert U_\alpha \vert \leq 1$.\\

\noindent Firstly, we show that for each simplex $\alpha \in K$, $\vert V_\alpha \vert \leq 1$. Consider a $0$-simplex $\alpha \in K$. Since, the dimension of a simplex cannot be less than $0$, for each $0$-simplex $\alpha \in K$ the corresponding set $V_\alpha$ is empty. In other words, for
each $0$-simplex $\alpha \in K$, we have that $\vert V_\alpha \vert = 0$. Also, for each $p$-simplex $\alpha^p \in K$ such that $1 \leq p \leq d$, Lemma 6 below shows that $\vert V_\alpha \vert \leq 1$. Thus, for every simplex $\alpha \in K$, we have shown that $\vert V_\alpha \vert \leq 1$.\\

\noindent Secondly, we show that for each simplex $\alpha \in K$, $\vert U_\alpha \vert \leq 1$. For each $p$-simplex $\alpha^p \in K$ such that $0 \leq p \leq (d-1)$, we prove in Lemma 5 below that $\vert U_\alpha \vert \leq 1$.  Now consider a $d$-simplex $\alpha^d \in K$. Since, by assumption $K$ is a $d$-dimensional simplicial complex, there are no $(d+1)$-simplices in $K$, and thus, the set $U_\alpha$ for each $d$-simplex $\alpha^d$ in $K$ is empty. In other words, for each $d$-simplex $\alpha^d \in K$, we have that $\vert U_\alpha \vert = 0$. Thus, for every simplex $\alpha \in K$ we have shown that $\vert U_\alpha \vert \leq 1$.\\

\noindent In summary, we have shown that for each simplex $\alpha \in K$, $\vert V_\alpha \vert \leq 1$ and $\vert U_\alpha \vert \leq 1$. Thus, $f$ satisfies the definition of a discrete Morse function on the simplicial complex $K$.\\
\qed

\noindent We next prove the  Lemmas used in the proof of the theorem above. This is done in the following sequence of Lemmas 2 to 6. We assume that $K$ is a $d$-dimensional simplicial complex and $f$ is the output function on $K$ obtained from algorithm 1.  We \textit{remark} that a $p$-dimensional simplex $\alpha$ of a simplicial complex $K$ is denoted by $\alpha^{p} \in K$. Also, if $p$-simplex $\alpha$ is a face of a $(p+1)$-simplex $\beta$ then this is represented as $\alpha^p < \beta^{p+1}$ in the sequel.\\

\noindent \textbf{Lemma 2.} \textit{For each $p$ where $0 \leq p \leq (d-1)$, if  $\alpha^{p} \in K$ and $\beta^{p+1} \in K$ such that $\alpha^{p} < \beta^{p+1}$, then, $f(\alpha) \neq f(\beta)$.}\\

\noindent \textit{Proof.} Let $\gamma_0,\gamma_1,\cdots,\gamma_{p+1}$ denote the $p$-dimensional faces of $\beta^{p+1}$ such that $f(\gamma_0) \geq f(\gamma_1) \geq \cdots \geq f(\gamma_{p+1})$. Note that $\alpha^{p}$ is one such $\gamma_i$ since by assumption $\alpha^{p}$ is a $p$-dimensional face of $\beta^{p+1}$. Based on lines 11-23 in algorithm 1, we have that:
\begin{equation}
\label{lem1eq}
f(\beta) =
\begin{cases}
(f(\gamma_0) + f(\gamma_1))/2 & \text{ if } \texttt{Flag}[\gamma_0] =0 \text{ and } f(\gamma_0) > f(\gamma_1)
\qquad \text{(Case $A$)}\\
f(\gamma_0) + \epsilon & \text{ otherwise} \qquad \qquad \qquad \qquad \qquad \qquad \; \text{(Case $B$)}
\end{cases}
\end{equation}
where $\epsilon > 0$.\\
\noindent Case $A$ implies $f(\gamma_0) >  f(\beta) > f(\gamma_1) \geq f(\gamma_2) \cdots \geq f(\gamma_{p})
\geq f(\gamma_{p+1})$.\\
\noindent Case $B$ implies $f(\beta) > f(\gamma_0) \geq f(\gamma_1) \geq f(\gamma_2) \cdots \geq f(\gamma_{p})
\geq f(\gamma_{p+1})$.\\
\noindent Thus, for both cases we have that $f(\beta) \neq f(\gamma_i)$ for each $i \in \{0,1,2,\cdots,(p+1)\}$.
Since $\alpha^{p}$ is one such $\gamma_i$ for some $i \in \{0,1,2,\cdots,(p+1)\}$, we have $f(\alpha) \neq f(\beta)$.\\
\qed

\noindent \textbf{Lemma 3.} \textit{For each $p$ where $0 \leq p \leq (d-1)$, if $\alpha^{p} \in K$ and $\beta^{p+1} \in K$ such that $\alpha^{p} < \beta^{p+1}$, then $f(\alpha) > f(\beta)$ if and only if} \texttt{Flag}$[\alpha]$ \textit{changes value from $0$ to $1$ while assigning function value for $\beta$.}\\

\noindent \textit{Proof.} Given $\alpha^{p} \in K$ and $\beta^{p+1} \in K$ \ such that \ $\alpha^{p} < \beta^{p+1}$, we first assume $f(\alpha) > f(\beta)$. Let $\gamma_0,\gamma_1,\cdots,\gamma_{p+1}$ denote the $p$-dimensional faces of $\beta^{p+1}$ such that $f(\gamma_0) \geq f(\gamma_1) \geq \cdots \geq f(\gamma_{p+1})$. Then, the value $f(\beta)$ is given by equation \ref{lem1eq}.\\
\noindent Case $A$ in equation \ref{lem1eq} implies $f(\gamma_0) >  f(\beta) > f(\gamma_1) \geq f(\gamma_2) \cdots \geq f(\gamma_{p}) \geq f(\gamma_{p+1})$.\\
\noindent Case $B$ in equation \ref{lem1eq} implies $f(\beta) > f(\gamma_0) \geq f(\gamma_1) \geq f(\gamma_2) \cdots \geq f(\gamma_{p}) \geq f(\gamma_{p+1})$. \\

\noindent Since by assumption, $\alpha^{p}$ is a face of $\beta^{p+1}$ and $f(\alpha) > f(\beta)$, Case $A$ is applicable, and we have $\gamma_0$ equals $\alpha^{p}$. Thus, based on line 18 in algorithm 1, while assigning the function value on $\beta^{p+1}$, \texttt{Flag}$[\alpha]$ changes value from $0$ to $1$.\\

\noindent Now, given $\alpha^{p} \in K$ and $\beta^{p+1} \in K$ \ such that \ $\alpha^{p} < \beta^{p+1}$, we assume that \texttt{Flag}$[\alpha]$ changes value from $0$ to $1$ while assigning function value for $\beta^{p+1}$. Let $\gamma_0,\gamma_1,\cdots,\gamma_{p+1}$ denote the $p$-dimensional faces of $\beta^{p+1}$ such that $f(\gamma_0) \geq f(\gamma_1) \cdots \geq f(\gamma_{p+1})$. Based on lines 11-23 in algorithm 1, \texttt{Flag}$[\alpha]$ changes
value from $0$ to $1$ while assigning function value on $\beta^{p+1}$ implies that $\gamma_0$ equals $\alpha^{p}$. Thus, we have that $f(\alpha) > f(\beta) = (f(\alpha) + f(\gamma_1))/2$.\\
\qed

\noindent \textbf{Lemma 4.} \textit{Let $\alpha$ be a simplex of $K$. Then, the number of times} \texttt{Flag}$[\alpha]$ \textit{changes value from $0$ to $1$ is $\leq 1$.}\\

\noindent \textit{Proof.} \texttt{Flag}$[\alpha]$ is initially set to $0$ in algorithm 1. In algorithm 1, if \texttt{Flag}$[\alpha]$ transitions to 1, its value never changes. In other words, there is no procedure in our algorithm 1 which changes the \texttt{Flag} of a simplex from 1 to 0. Thus, either \texttt{Flag}$[\alpha]$ remains 0 throughout or changes value from 0 to 1 exactly once in algorithm 1.\\
\qed

\noindent \textbf{Lemma 5.} \textit{For each $p$ where $0 \leq p \leq (d-1)$, if $\alpha^{p} \in K$, then $ \vert U_\alpha \vert \leq 1$.}\\

\noindent \textit{Proof.} From Lemma 3, we have that for each $\alpha^{p} \in K$, the number of $(p+1)$-simplices $\beta^{p+1}$ such that $\alpha^{p} < \beta^{p+1}$ and $f(\beta) < f(\alpha)$ is equal to the number of times \texttt{Flag}$[\alpha]$ changes value from $0$ to $1$. Applying Lemma 4, we get that for each $\alpha^{p} \in K$,
\begin{equation}
\nonumber
\# \{ \beta^{p+1} \: | \: \alpha^{p} < \beta^{p+1} \text{ and } f(\beta) < f(\alpha) \} \leq 1.
\end{equation}
\noindent Furthermore, Lemma 2 tells us that,
\begin{equation}
\nonumber
\# \{ \beta^{p+1} \: | \: \alpha^{p} < \beta^{p+1} \text{ and } f(\beta) < f(\alpha) \} =  \# \{ \beta^{p+1} \: 
| \: \alpha^{p} < \beta^{p+1} \text{ and } f(\beta) \leq f(\alpha) \}
\end{equation}
\noindent Thus, we have that for each $\alpha^{p} \in K,\ \vert U_{\alpha} \vert$ = \#$\{ \beta^{p+1} \: | \:
\alpha^{p} < \beta^{p+1} \text{ and } f(\beta) \leq f(\alpha) \} \; \leq \;1$.\\
\qed

\noindent \textbf{Lemma 6.} \textit{For each $p$ \textit{where} $1 \leq p \leq d$, if $\alpha^p \in K$, then $\vert V_\alpha \vert \leq 1$.}\\

\noindent \textit{Proof.} Let $\gamma_0,\gamma_1,\cdots,\gamma_{p}$ denote the $(p-1)$-dimensional faces of $\alpha^p \in K$ such that $f(\gamma_0) \geq f(\gamma_1) \geq \cdots \geq f(\gamma_{p})$. Based on lines 11-23 in algorithm 1, we have that:
\begin{equation}
\nonumber
f(\alpha) =
\begin{cases}
(f(\gamma_0) + f(\gamma_1))/2 & \text{ if } \texttt{Flag}[\gamma_0] =0 \text{ and } f(\gamma_0) > f(\gamma_1)
\qquad \text{(Case $A$)}\\
f(\gamma_0) + \epsilon & \text{ otherwise} \qquad \qquad \qquad \qquad \qquad \qquad \; \text{(Case $B$)}
\end{cases}
\end{equation}
where $\epsilon > 0$.\\
\noindent Case $A$ implies  $f(\gamma_0) >  f(\alpha) > f(\gamma_1) \geq f(\gamma_2) .... \geq f(\gamma_{p-1}) \geq f(\gamma_{p}) $, and thus, $\vert V_\alpha \vert = 1$.\\
\noindent Case $B$ implies $f(\alpha) > f(\gamma_0) \geq f(\gamma_1) \geq f(\gamma_2) .... \geq f(\gamma_{p-1}) \geq f(\gamma_{p})$, and thus, $\vert V_\alpha \vert = 0$.\\
\noindent Hence, for each $\alpha^p \in K$ with $1 \leq p \leq d$, we have that $\vert V_\alpha \vert \leq 1$.\\
\qed


\section*{Filtration of the clique complex based on weights of critical simplices}

Given a simplicial complex $K$, its dimension $d$ and a discrete Morse function $f$ on $K$, the algorithm 2 determines the weights of critical simplices in $K$. In the pseudocode of the algorithm 2, lines 2-6 initialize a variable \texttt{IsCritical}$[\alpha]$ associated to every simplex $\alpha$ in clique complex $K$ to be True. Lines 7-17 determine the critical simplices in $K$ by checking for the condition in equation 7 of the main text which defines a critical simplex. Lines 18-28 determine the weights of critical simplices or critical weights in $K$. Finally, the algorithm 2 outputs an array $w_c[ ]$ which contains an increasing sequence of critical weights in $K$. Subsequently, this increasing sequence of critical weights will be used for the filtration of the clique complex $K$.

Given an unweighted and undirected graph $G$, we restrict the construction of clique complex $K$ by including simplices up to a maximum dimension $d$. Then, the algorithm 3 creates the filtration of clique complex $K$ based on weights of critical simplices as described in the Theory section of the main text. In the pseudocode of the algorithm 3, lines 2-6 assigns a non-negative function $g$ to $0$-simplices in clique complex $K$. Line 7 calls the algorithm 1 for the assignment of weights satisfying discrete Morse function to every simplex in $K$. Line 8 calls the algorithm 2 to obtain an increasing sequence of unique weights corresponding to critical simplices in $K$. Lines 9-11 initialize a variable \texttt{IsAdded}$[\alpha]$ associated to every simplex $\alpha$ in $K$ which tracks whether the simplex $\alpha$ has been added to the filtration or not. Lines 12-31 compute the filtration weight of each simplex $\alpha$ in $K$ as described in the Theory section of the main text. 

In SI Table S1, we describe the role of key variables which appear in algorithms 2 and 3.

\begin{figure}
\setcounter{algorithm}{1}
\begin{algorithm}[H]
\caption{Algorithm to compute the weights of critical simplices in $K$ corresponding to the discrete Morse function $f$}
\begin{algorithmic}[1]
\Function{GetCriticalWeights}{$K,d,f$}
\Statex
    \For{$p=0,\cdots,d$}                   \Comment{Initialize \texttt{IsCritical} variable associated with each simplex in $K$}
	\For{each $p$-simplex $\alpha \in K$} 	
	   \State \texttt{IsCritical}$[\alpha]$= True \Comment{\texttt{IsCritical} indicates whether a given simplex is critical or not}
	\EndFor
    \EndFor
    \Statex
	\For{$p=1,\cdots,d$}                   \Comment{Determine the critical simplices in $K$}
	\For{each $p$-simplex $\alpha \in K$}
		\State Let \texttt{Faces}[ ] be an array of all $(p-1)$-dimensional faces of $\alpha$
		\State Sort \texttt{Faces}[ ] such that $f(\texttt{Faces}[i]) \geq f(\texttt{Faces}[i+1])$ for each $i \in \{0,1,\cdots,p-1\}$
		\State Let $\gamma_0 = \texttt{Faces}[0]$
		\If{$f(\gamma_0)\geq f(\alpha)$ }
				\State \texttt{IsCritical}[$\alpha] = $ False
				\State \texttt{IsCritical}$[\gamma_0] = $ False
		\EndIf
	\EndFor
	\EndFor
	\Statex
    \State Initialize $i = 0	$
	\State Declare empty array $w_c$[ ]
    \For{$p=0,\cdots,d$}                    \Comment{Determine the weights of critical simplices in $K$}
	\For{each $p$-simplex $\alpha \in K$}
		\If{ \texttt{IsCritical}$[\alpha]$ = True}	
			\State $w_c[i] = f(\alpha)$
			\State $i = i+1$
		\EndIf
	\EndFor
    \EndFor
	\State Sort array $w_c$[ ] in increasing order and remove duplicates
    \Statex
	\State \Return $w_c$[ ]		
\Statex
\EndFunction
\end{algorithmic}
\end{algorithm}	
\end{figure}

\begin{figure}
\begin{algorithm}[H]
\caption{Algorithm to determine the filtration weights of the simplices in the clique complex of a simple graph $G$ }
\begin{algorithmic}[1]
    \Statex
	\State $K$ = $d$-dimensional clique complex of graph $G$
	\Statex
    \State $deg_{max} =$ Maximum degree of a vertex in graph $G$
    \For{ each $0$-simplex $\alpha \in$ $K$ }                        \Comment{Assign non-negative function $g$ to $0$-simplices in $K$}
		\State $\epsilon =$ random$(0,0.5)$ 					\Comment{$\epsilon$ is  generated randomly during runtime using a uniform distribution on $(0,0.5)$}
		\State $g[\alpha]$ = $deg_{max}$ - degree($\alpha$) + $\epsilon$
	\EndFor	
	\Statex
	\State $f$ = \textproc{DisceteMorseFunction($K,d,g$)}	         \Comment{Call Algorithm 1}
	\State $w_c$[ ] = \textproc{GetCriticalWeights($K,d,f$)}           \Comment{Call Algorithm 2}
	\Statex
	\For{each simplex $\alpha \in K$} 	                             \Comment{Initialize \texttt{IsAdded} variable associated with each simplex in $K$}
		\State \texttt{IsAdded}$[\alpha]= False$ \Comment{\texttt{IsAdded} indicates whether a given simplex has been added to the filtration.}
	\EndFor
	\Statex
	\For{$i=0,\cdots,len(w_c$[ ]$)-1$} 	                              \Comment{Calculate Filtration weight for each simplex in $K$}
		\For{each simplex $\alpha \in K$} 	
			\If{ $f(\alpha) \leq w_c[i]$  AND  \texttt{IsAdded}$[\alpha] = False$ }
				\State \texttt{FiltrationWeight}$[\alpha] = w_c[i]$ 					
				\State \texttt{IsAdded}$[\alpha] = True$
				\For{each face $\gamma < \alpha $}
					\If{	\texttt{IsAdded}$[\gamma] = False$}
						\State \texttt{FiltrationWeight}[$\gamma$] = $w_c[i]$
						\State \texttt{IsAdded}$[\gamma] = True$
					\EndIf
				\EndFor
			\EndIf
		\EndFor
	\EndFor	
    \Statex

    \For{each simplex $\alpha \in K$}
		\If{ \texttt{IsAdded}$[\alpha] = False$ }
    		\State \texttt{FiltrationWeight}$[\alpha] = w_c[len(w_c$[ ]$)-1]$
    		\State \texttt{IsAdded}$[\alpha] = True$
    	\EndIf	
    \EndFor
\end{algorithmic}
\end{algorithm}
\end{figure}

\section*{Computational aspects of our algorithms}

\subsection*{Time complexity of Algorithm 1}

We briefly discuss here some computational aspects of the algorithm 1 in the main text. Given a simplicial complex $K$, its dimension $d$ and a non-negative real-valued function $g$ on the 0-simplices of $K$, algorithm 1 assigns weights to any simplex in $K$, producing a discrete Morse
function $f$ on $K$.

We remark that the algorithms 1, 2 and 3 were implemented using the C++ programming language. The simplices of a $d$-dimensional simplicial complex were stored using the set container which is a part of the C++ Standard Template Library (STL). The association of the discrete Morse function values to the corresponding simplices was implemented using the map container of the C++ STL. Implementing the storage of the \texttt{Flag} variable associated with each simplex was also done using the map container of C++ STL. We remark that accessing a particular value of a map container has a complexity which is logarithmic with respect to the size of the container, namely, the number of simplices $n$.

Assuming that the dimension $d$ of $K$ is constant, we here present a brief discussion about the complexity of algorithm 1 with respect to the number of simplices $n$ in $K$. We remark that since the dimension $d$ is constant and the complexity of accessing values using map containers is logarithmic with respect to $n$, lines 2-9 in algorithm 1 have a complexity of $O(n\log{}n)$. The complexity of the operation corresponding to line 12 in algorithm 1, namely, finding the $(p-1)$-dimensional faces of a $p$-simplex is independent of $n$. Such an operation depends on $p$ which has an upper bound of $d$ that has been assumed to be a constant.

The operation in line 13 of algorithm 1 corresponds to arranging the simplices computed in the preceding line based on their function values. As noted earlier, accessing values using map containers is logarithmic with respect to $n$. Thus, since $p \leq d$ and $d$ is fixed, the operation represented
by line 13 of algorithm 1 has a complexity of $O(\log{}n)$. Lines 14-22 of algorithm 1 except the random number generation in line 20 of algorithm 1 require accessing values of map containers. Random number generation based on a uniform distribution using C++ can be implemented with amortized constant
complexity. Thus, the block of pseudocode represented by lines 14-22 in algorithm 1 have complexity of $O(\log{}n)$. Since $d$ is constant and for any $p$ the number of $p$-simplices is bounded by the total number of simplices $n$, the pseudocode represented in lines 10-24 of algorithm 1 has a complexity of
$O(n\log{}n)$. Hence the complexity of algorithm 1 with respect to the number of simplices $n$ can be described as $O(n\log{}n)$.


\subsection*{Time complexity of algorithms 1, 2 and 3 with respect to the number of critical simplices}

Firstly, it is apparent that the complexity of algorithm 1 is independent of the number of critical simplices since the notion of critical simplices is relevant only after the discrete Morse function has been constructed.

Algorithm 2 accepts as inputs a simplicial complex $K$, its dimension $d$ and a discrete Morse function $f$ on $K$ and outputs the sorted sequence of weights which correspond to the critical simplices in $K$. Let $m$ denote the total number of critical simplices in $K$ corresponding to $f$. We assume henceforth that the simplicial complex $K$ and its dimension $d$ are fixed. This implies that the number of simplices $n$ and the dimension $d$ are treated as constants. In such a situation, the complexity of all lines in the pseudocode of algorithm 2 except for line 28 is constant and independent of $m$. Line 28 of algorithm 2 represents sorting the array which contains the function values of all critical simplices, i.e., the critical weights, and removing any duplicate entries. The number of critical weights is less than or equal to the number of critical simplices $m$, and thus, the operation represented by line 28 of algorithm 2 can be implemented with a complexity of $O(m^2)$. Hence, assuming that the simplicial complex $K$ and its dimension $d$ are fixed, algorithm 2 can be implemented with a complexity of $O(m^2)$ in respect to the number of critical simplices $m$.

Algorithm 3 accepts a graph as input and determines the filtration weights of the simplices in the $d$-dimensional clique complex $K$ of $G$. Similar to the analysis of algorithm 2, we assume that the graph $G$ and the dimension $d$ of $K$ is fixed which implies that the number of simplices in the $d$-dimensional clique complex $K$ of $G$ is also constant. In such a situation, lines 1-7 of algorithm 3 have constant complexity which is independent of the number of critical simplices $m$. As seen in the preceding paragraph, provided $K$ and $d$ are fixed, the complexity of algorithm 2 and hence the complexity of line 8 of algorithm 3 is $O(m^2)$. Lines 9-11, lines 13-24 and lines 26-31 in the pseudocode of algorithm 3 have constant complexity which is independent of the number of critical simplices $m$. The number of times the operations of lines 13-24 in algorithm 3 are repeated is equal to the number of critical weights which is less than or equal to the number of critical simplices $m$. Thus, assuming that the graph $G$ and the dimension $d$ of the corresponding clique complex $K$ are fixed, algorithm 3 can be implemented with a complexity of $O(m^2)$ in respect to the number of critical simplices $m$.


\section*{Theoretical results on stability of persistent homology and persistence diagrams of discrete
Morse functions}

\subsection*{Stability of persistent homology}

In Algorithm 1, some choices are made in assigning values to every simplex in order to get a discrete Morse function. A natural question arises which is the following. Under what conditions do two discrete Morse functions give the same persistent homology groups. We remark that in Forman's theory \cite{Forman2002}, the actual values of the discrete Morse functions are less important than the \textit{gradient vector field} that the discrete Morse function induces. For the definition of a gradient vector field, we refer the reader to Definition 3.3 in Forman's \cite{Forman2002} article. In his expository paper \cite{Forman2002}, Forman asserts on page 15 that: \textit{``In fact, this gradient vector field contains all of the information that
we will need to know about the function for most applications.''}

Note that two Morse functions $f$ and $g$ induce the same gradient vector field if and only if they satisfy the following condition (see Forman \cite{Forman2005} or Theorem 3.1 in Ayala \textit{et al.}\cite{Ayala2009}):
\begin{equation}
\nonumber
\textbf{C0}: f(\alpha)<f(\beta) \Leftrightarrow g(\alpha)< g(\beta) \text{ for a $p$-simplex } \alpha
\text{ and a $(p+1)$-simplex } \beta \text{ such that } \alpha^p< \beta^{p+1}
\end{equation}

We show that for two discrete Morse functions to have the same filtration given by their level subcomplexes on critical weights, the actual values of the discrete Morse function are not as important as the \textit{order relation} induced by the function on the simplices. The discrete Morse functions with the same filtration clearly give the same persistent homology groups. In particular, we show below that a sufficient condition for this to happen is that both functions have the same order relation with respect to the values assigned to the simplices.

\vspace{5pt}
\noindent \textbf{Lemma 7.}
\textit{Let $f$ and $g$ be two discrete Morse functions on a finite simplicial complex $K$. Assume that the following condition is satisfied:}
\begin{equation}
\nonumber
\textbf{C1}: f(\alpha) \leq f(\beta) \Leftrightarrow g(\alpha)\leq g(\beta) \text{ for any pair of
simplices } \alpha, \beta \in K.
\end{equation}
\textit{Then $f$ and $g$ have the same set of critical simplices $C:=\{ c_1, c_2, \cdots, c_k\}$, and the level subcomplex of $f$ corresponding to the critical cell $c_i$ is the same as the level subcomplex of $g$ corresponding to $c_i$, for any $i= 1, 2, \cdots, k.$}

\vspace{5pt}
\noindent \textit{Proof.} The condition \textbf{C1} guarantees that the functions $f$ and $g$ have the same gradient vector field, since \textbf{C1} implies \textbf{C0}. Therefore they have the same set of critical simplices $C$. Denote the level subcomplex of $f$ (respectively, of $g$) corresponding to the critical cell $c_i$ by $K^f(c_i)$ (respectively, by $K^g(c_i)$). Now, if $\alpha \in K^f(c_i)$, then there exists $\beta \in K$ such that $\alpha \leq \beta$ and $f(\beta)\leq f(c_i)$. Note that $\alpha \leq \beta$ represents that either two simplices $\alpha$ and $\beta$ are same or $\alpha$ is a face of $\beta$. By assumption \textbf{C1} this implies that $g(\beta)\leq g(c_i)$, thus $\alpha\in K^g(c_i)$. So we get $K^f(c_i)\subseteq K^g(c_i)$. Reversing the roles of $f$ and $g$ in the above argument, we also get $K^g(c_i)\subseteq K^f(c_i)$. Thus $K^f(c_i)= K^g(c_i)$ for any critical cell $c_i \in C$.
\qed
\vspace{5pt}

A simple example of two discrete Morse functions $f$ and $g$ which satisfy \textbf{C1} is the following. Starting with a discrete Morse function $f$ obtain $g$ by \textit{sliding} the values of $f$ by a fixed real number $r$. A slightly more sophisticated example of two functions which satisfy \textbf{C1} is the following. Given a discrete Morse function $f$ on a finite simplicial complex $K$, one can arrange the values of $f$ on the simplices of $K$ in non-decreasing order. Suppose that $\alpha$ and $\beta$ are simplices of $K$ which are adjacent in this ordering, i.e., we have $f(\alpha)\leq f(\beta)$ and no other simplex $\sigma$ exists such that $f(\sigma)$ lies in the interval $[f(\alpha), f(\beta)]$, and let us call such pairs of simplices $(\alpha, \beta)$ as \textit{$f$-successive}. If for each $f$-successive pair of simplices $(\alpha, \beta)$, one adds a non-negative number $\delta(\alpha, \beta)$ to $f(\alpha)$ such that $\delta(\alpha, \beta) \leq f(\beta)- f(\alpha)$, we obtain a new discrete Morse function $g$ given by
$g(\alpha):= f(\alpha)+ \delta(\alpha, \beta)$, for which any pair $(\alpha,\beta)$ is $f$-successive if and only if it is $g$-successive. It is easy to deduce that $f$ and $g$ then satisfy the condition \textbf{C1}. Thus, $f$ and $g$ will have the same persistent homology groups.


\subsection*{Stability of persistence diagrams}

We remark that two discrete Morse functions may have the same filtration (and thus the same persistent homology groups) but their persistence diagrams may differ if their critical weights (i.e., values on the critical simplices) are different. A sufficient condition that assures that two discrete Morse functions $f$ and $g$ have the same persistence diagrams is the following, keeping the notations of Lemma 7:
\begin{equation}
\nonumber
\textbf{C2:} \text{ in addition to \textbf{C1}, we have } f(c_i)=g(c_i) \text{ for any } c_i \in C
\end{equation}

Even if the condition \textbf{C2} is not satisfied, there are stability theorems which give conditions under which the bottleneck distance between the respective persistence diagrams does not change much, i.e., the persistence diagrams are \textit{close} in the bottleneck metric. The first such stability theorem was given by Cohen-Steiner \textit{et al.} \cite{Cohen-Steiner2007}. This result \cite{Cohen-Steiner2007} is not directly applicable to discrete
Morse functions, as they are not continuous. However, we show below that stability theorems for discrete Morse functions for finite regular CW-complexes can be inferred from stability results in Chazal \textit{et al.} \cite{Chazal2008} (see also Bauer \textit{et al.}\cite{Bauer2012}). For simplicity we give the arguments for discrete Morse functions on a finite simplicial complex, although all the results below are also valid for finite regular
CW-complexes.

Recall the $\infty$-Wasserstein distance or bottleneck distance between two multisets $X$ and $Y$ in $\mathbb{R}^2$:
\begin{equation}
\label{botdist}
W_\infty(X,Y) =  \inf_{\eta:X \rightarrow Y} \text{sup}_{x \in X} || x - \eta(x)
||_{\infty}.
\end{equation}
In the above equation, the supremum is taken over all bijections $\eta:X\rightarrow Y$ (with the convention that a point with multiplicity $k \in \N$ is considered as $k$ individual points) and for $(a,b) \in \mathbb{R}^2$, $||(a,b)||_\infty:= \max\{|a|, |b|\}$.

For a discrete Morse function $f$ with critical values $w_i:=f(c_i), i= 1, 2, \cdots, n$, the $k^{th}$ persistence diagram $D^kf$ (or simply $Df$ when the value of $k$ is fixed), is defined as follows. Consider the multiset of points $W^k_f:=\{(w_i, w_j): w_i<w_j, i, j = 1, 2, \cdots, n\}$ with each point $(w_i, w_j)$ endowed with the multiplicity $\mu_k(w_i, w_j)$ given by (see e.g. Di Fabio \textit{et al.} \cite{DiFabio2015}):
\begin{equation}
\nonumber
\mu_k(w_i, w_j) := \lim_{\epsilon \rightarrow 0^+} (\beta_{w_i+\epsilon}^{w_j-\epsilon} - \beta_{w_i+\epsilon}^{w_j+\epsilon}+\beta_{w_i-\epsilon}^{w_j+\epsilon}- \beta_{w_i-\epsilon}^{w_j-\epsilon})
\end{equation}
where $\beta_x^y:= \rank(H_k(K^f(x))\rightarrow H_k(K^f(y)))$ for $x, y \in \R$ with $x<y$. Denote by $\Delta$ the diagonal in $\R^2$ considered as a multiset with infinite multiplicity given to each of its points.

\vspace{5pt}
\noindent \textbf{Definition 8.}
\textit{The persistence diagram $D^kf$ is the subset of $W^k_f\cup \Delta$ consisting of points $(u, v)$ with $\mu_k(u, v)>0$.}
\vspace{5pt}

\noindent We state below the stability results from Chazal \textit{et al.} \cite{Chazal2008} that we shall need, and to do so,
we recall the concepts of \textit{persistence modules} and $\epsilon$-\textit{interleaving}:

\vspace{5pt}
\noindent \textbf{Definition 9.} (Persistence modules (Definition 2.2. in Chazal \textit{et al.} \cite{Chazal2008})) \textit{Let $R$ be a commutative ring with unity. A collection $\maF=\{F^\alpha\}_{\alpha \in \mathbb{R}}$ of $R$-modules $F^\alpha$ together with homomorphisms $\phi_{\alpha\beta}: F^\alpha\rightarrow F^\beta$ for $\alpha\leq \beta$, is called a persistence module if}
\begin{equation}
\nonumber
\forall \alpha\leq \beta\leq \gamma, \quad \quad \phi_{\alpha\alpha}= \id_{F^\alpha} \text{ and } \quad \quad
\phi_{\alpha\beta}\circ \phi_{\beta\gamma}=\phi_{\alpha\gamma}
\end{equation}
\textit{A persistence module is called $\delta$-tame if for any $\alpha < \alpha+\delta <\beta$ one has $\rank \phi_{\alpha\beta} <\infty$.}
\vspace{5pt}

\vspace{5pt}
\noindent \textbf{Definition 10.} ($\epsilon$-interleaving of functions (Definition 4.1 in Chazal \textit{et al.}
\cite{Chazal2008}))
\textit{Let $\epsilon>0$, and let $f$ and $g$ be two real-valued functions on a finite simplicial complex $K$. For $x \in \R$, denote the sub-level sets $f^{-1}(-\infty, x]$ of $f$ (respectively, $g$) by $F^x$ (respectively, $G^x$). The functions $f$ and $g$ are said to be strongly $\epsilon$-interleaved if for any $a\in \R$ and $n \in \Z$, one has}
\begin{equation}
\nonumber
F^{a+2n\epsilon} \subseteq G^{a+(2n+1)\epsilon} \subseteq F^{a+(2n+2)\epsilon}
\end{equation}
\vspace{5pt}

\noindent The main result for $\epsilon$-interleaved functions that we shall need is as follows:

\vspace{5pt}
\noindent \textbf{Lemma 11.} (Lemma 4.2(iii) in Chazal \textit{et al.} \cite{Chazal2008})
\textit{Given $\epsilon>0$, two functions $f$ and $g$ are strongly $\epsilon$-interleaved if and only if $||f-g||_\infty < \epsilon$.}
\vspace{5pt}

\vspace{5pt}
\noindent \textbf{Definition 12.} ($\epsilon$-interleaving of persistence modules (Definition 4.3 in Chazal \textit{et al.} \cite{Chazal2008}))
\textit{Let $\maF=\{F^\alpha\}_{\alpha \in \R}$ and $\maG=\{G^\alpha\}_{\alpha\in \R}$ be persistence modules over a commutative unital ring $R$. Let $\epsilon>0$, then $\maF$ and $\maG$ are said to be strongly $\epsilon$-interleaved if for all $a\in \R$ and $n\in \Z$, there exist homomorphisms
$\Phi_{a+2n\epsilon}: F^{a+2n\epsilon} \rightarrow G^{a+(2n+1)\epsilon}$ and $\Psi_{a+(2n+1)\epsilon}: G^{a+(2n+1)\epsilon} \rightarrow F^{a+(2n+2)\epsilon}$ such that the following diagram commutes:}
\[
\begin{tikzcd}[column sep=huge, row sep=huge]
     \cdots \arrow[r] & F^{a+2n\epsilon} \arrow[r]  \arrow[rd] &  F^{a+(2n+1)\epsilon} \arrow[r]  & F^{a+(2n+2)\epsilon} \arrow[r] \arrow[rd]  & \cdots \\
     \cdots \arrow[r] \arrow[ru] & G^{a+2n\epsilon} \arrow[r]  &  G^{a+(2n+1)\epsilon} \arrow[r] \arrow[ru] & G^{a+(2n+2)\epsilon}  \arrow[r]  & \cdots
\end{tikzcd}
\]
\vspace{5pt}

If $f$ and $g$ are strongly $\epsilon$-interleaved, then the associated persistence modules given by their $k^{th}$ singular homology groups are also strongly $\epsilon$-interleaved for any $k \in \N \cup \{0\}$, the associated homomorphisms in Definition 10 are induced by the corresponding inclusion maps. Note that these persistence modules are $0$-tame since the functions are defined on a finite simplicial complex $K$.

One can define persistence diagrams for $\delta$-tame persistence modules (see Definitions 3.6 and 3.10 in Chazal \textit{et al.} \cite{Chazal2008}). Denote the persistence diagram of a $\delta$-tame persistence module $\maF$ by $D_\delta \maF$. The persistence diagram of persistence modules generalizes the classical persistence diagrams associated with real-valued functions, i.e., if $f$ is a tame real-valued function on a finite simplicial complex $K$, the associated $0$-tame persistence module given by the $k^{th}$ singular homology groups of its sublevel sets $F^\alpha$, denoted $\maF_k:= \{ H_k(F^\alpha)\}_{\alpha\in \R}$, has the same persistence diagram as the $k^{th}$ persistence diagram of $f$, i.e. $D^kf = D_0\maF_k$ (see Remark 1, page 12 in Chazal \textit{et al.} \cite{Chazal2008} for a short proof).

The next lemma shows that if $f$ is a discrete Morse function and $K^f(\alpha), \alpha\in \R$ are the level subcomplexes of $f$, then the persistence diagram of the $0$-tame persistence module $\maK_k= \{ H_k(K^f(\alpha))\}_{\alpha\in \R}$ is the same as the $k^{th}$ persistence diagram $D^kf$ of $f$
corresponding to the filtration of $K$ given by the level subcomplexes at critical weights, i.e., $D^kf= D_0\maK_k$. The proof is essentially the same as the arguments given in Remark 1, page 12 of Chazal \textit{et al.} \cite{Chazal2008}, except that we use Forman's Lemma 2.6 \cite{Forman2002} for comparing homology groups across critical weights. Note that due to Forman's Lemma 2.6 \cite{Forman2002}, the set of $k$-homological critical values of $f$, i.e., points $c \in \R$ such that for any $\epsilon>0$ the homomorphism on the $k$-homology groups induced by inclusion map $K^f(c-\epsilon)\hookrightarrow K^f(c+\epsilon)$ is \textit{not} an isomorphism, is a subset of its critical weights in general. However, it is easy to see that only those points $(w_i, w_j)$ appear in the persistence diagram of $f$ for which both $w_i$ and $w_j$ are $k$-homological critical values of $f$ (otherwise the multiplicity of such points is zero).

\vspace{5pt}
\noindent \textbf{Lemma 13.} \textit{We have $D^kf= D_0\maK_k$.}
\vspace{5pt}

\noindent \textit{Proof.} Let $k$ be fixed, to simplify the notation we denote the $k$-persistence diagrams of $f$ and $\maK_k$ by $Df$ and $D\maK$, respectively. From the definition of the persistence diagrams of persistence modules (Definition 3.6 in Chazal \textit{et al.} \cite{Chazal2008}), it suffices to show that there exists $\epsilon>0$ $x\in (0, \epsilon)$, such that $W_\infty(Df, D\maK_{\frac{\epsilon}{2^n}, x})<\epsilon/2^n$ for any $n\geq 1$, where $D\maK_{\frac{\epsilon}{2^n}, x}$ is the multi-subset of $\R^2$ given by points of non-zero multiplicity in the union of the grid
\begin{equation}
\nonumber
G_{\epsilon/2^n, x} := \{ (x+\frac{m\epsilon}{2^n}, x+\frac{m'\epsilon}{2^n}) \in \R^2, m, m' \in \Z: m'> m\}
\end{equation}
and the diagonal $\Delta$, where the multiplicity
$\mu(m,m')$ of a point $(x+\frac{m\epsilon}{2^n}, x+\frac{m'\epsilon}{2^n}) \in G_{\epsilon/2^n, x}$ is given by the following formula from Definition 3.1(ii) in Chazal \textit{et al.} \cite{Chazal2008}:
\begin{eqnarray*}
\mu(m, m')&:=& \rank(H(m)\rightarrow H(m'-1)) - \rank(H(m)\rightarrow H(m'))\\
&+&  \rank(H(m-1)\rightarrow H(m')) - \rank(H(m-1)\rightarrow H(m'-1))
\end{eqnarray*}
where we have used the notation $H(m):= H_k(K^f(x+\frac{m\epsilon}{2^n})), m\in \Z$ for simplicity.

Since the number of critical weights is finite, one can choose $\epsilon>0$, $x\in (0, \epsilon]$ such that none of the grids $G_{\epsilon/2^n, x}$ have a point outside the diagonal that coincides with a point in $Df$. Then, for each $n \geq 1$ a point $(w_i, w_j) \in Df$ is contained in exactly one of the cells $C$ in the grid $G_{\epsilon/2^n, x}$. It is easy to check using Forman's Lemma 2.6 \cite{Forman2002} that the multiplicity $\mu(w_i, w_j)$ is then equal to the multiplicity of the upper-right corner of $C$, while all other off diagonal points in $D\maK$ which are not upper-right corners of cells containing a point of $Df$, have multiplicity zero. Therefore the bijection $\Phi_{\epsilon,n}: Df\rightarrow D\maK_{\epsilon/2^n, x}$ which maps $(w_i, w_j)$ in $Df$ to the upper right corner of the unique cell in $G_{\epsilon/2^n, x}$ containing $(w_i, w_j)$ and any point in the region $\{(x,y) \in \R^2: y\in [x, x+\epsilon/2^n]\}$ to the nearest point on the diagonal, moves points by a distance at most $\frac{\epsilon}{2^n}$. Thus, we get $W_\infty(Df, D\maK_{\frac{\epsilon}{2^n}, x})<\epsilon/2^n$ for any $n\geq 1$.
\qed
\vspace{5pt}

\noindent We can now state the stability theorem for $\epsilon$-interleaved persistence modules:

\vspace{5pt}
\noindent \textbf{Theorem 14.} (Theorem 4.4 in Chazal \textit{et al.} \cite{Chazal2008})
\textit{Let $\maF$ and $\maG$ be two strongly $\epsilon$-interleaved persistence modules which are $\delta$-tame for some $\delta\geq 0$. Then,}
\begin{equation}
\nonumber
W_\infty(D_\delta \maF, D_\delta \maG) < 3\epsilon.
\end{equation}
\vspace{5pt}

\noindent We are now ready to state a stability theorem for persistence diagrams for discrete Morse functions:

\vspace{5pt}
\noindent \textbf{Theorem 15.} \textit{Let $f$ and $g$ be discrete Morse functions on a finite simplicial complex $K$ which induce $k^{th}$ persistence diagrams $D^kf$ and $D^kg$, respectively. Suppose that there exists $\epsilon>0$, such that $||f-g||_\infty <\epsilon$. Then,}
\begin{equation}
\nonumber
W_\infty(D^kf, D^kg) < 3\epsilon.
\end{equation}

\vspace{5pt}
\noindent \textit{Proof.} Let us show that the level subcomplexes of $f$ and $g$ are strongly $\epsilon$-interleaved. Indeed, since $||f-g||_\infty<\epsilon$, $f$ and $g$ are strongly $\epsilon$-interleaved by Lemma 11. Let $c \in \R$ and let $\sigma \in K^f(c)$ and denote by $F^c$ the sub-level set $f^{-1}((-\infty, c])$ of $f$; $G^c$ is defined similarly for the discrete Morse function $g$. Then, by Lemma 3.2 of Forman \cite{Forman1998}, there exists a co-face $\tau$ of $\sigma$ such that $f(\tau)\leq c$, i.e. $\tau \in F^c$. By the strong $\epsilon$-interleaving property, one gets $F^c \subseteq G^{c+\epsilon}$, so we have $g(\tau)\leq c+\epsilon$, which implies that $\sigma \in K^g(c+\epsilon)$, so that we get $K^f(c)\subseteq K^g(c+\epsilon)$. Similarly one easily gets $K^g(c)\subseteq K^f(c+\epsilon)$ for any $c \in \R$. This implies that the persistence modules of $f$ and $g$ given by the $k^{th}$ homology groups of the level subcomplexes are strongly $\epsilon$-interleaved for any $k\in \N\cup \{0\}$. Hence by Theorem 14, one gets:
\begin{equation}
\nonumber
W_\infty(D^kf, D^kg) < 3\epsilon
\end{equation}
This concludes the proof.
\qed
\vspace{5pt}

As a concluding remark to this section, we observe that the bound of $3\epsilon$ can be improved to $\epsilon$, following the same arguments as given in the proof of Theorem 4.8 in Chazal \textit{et al.} \cite{Chazal2008}, we omit the details here.


\setcounter{figure}{0}
\begin{figure*}
\centering
\includegraphics[width=.9\columnwidth]{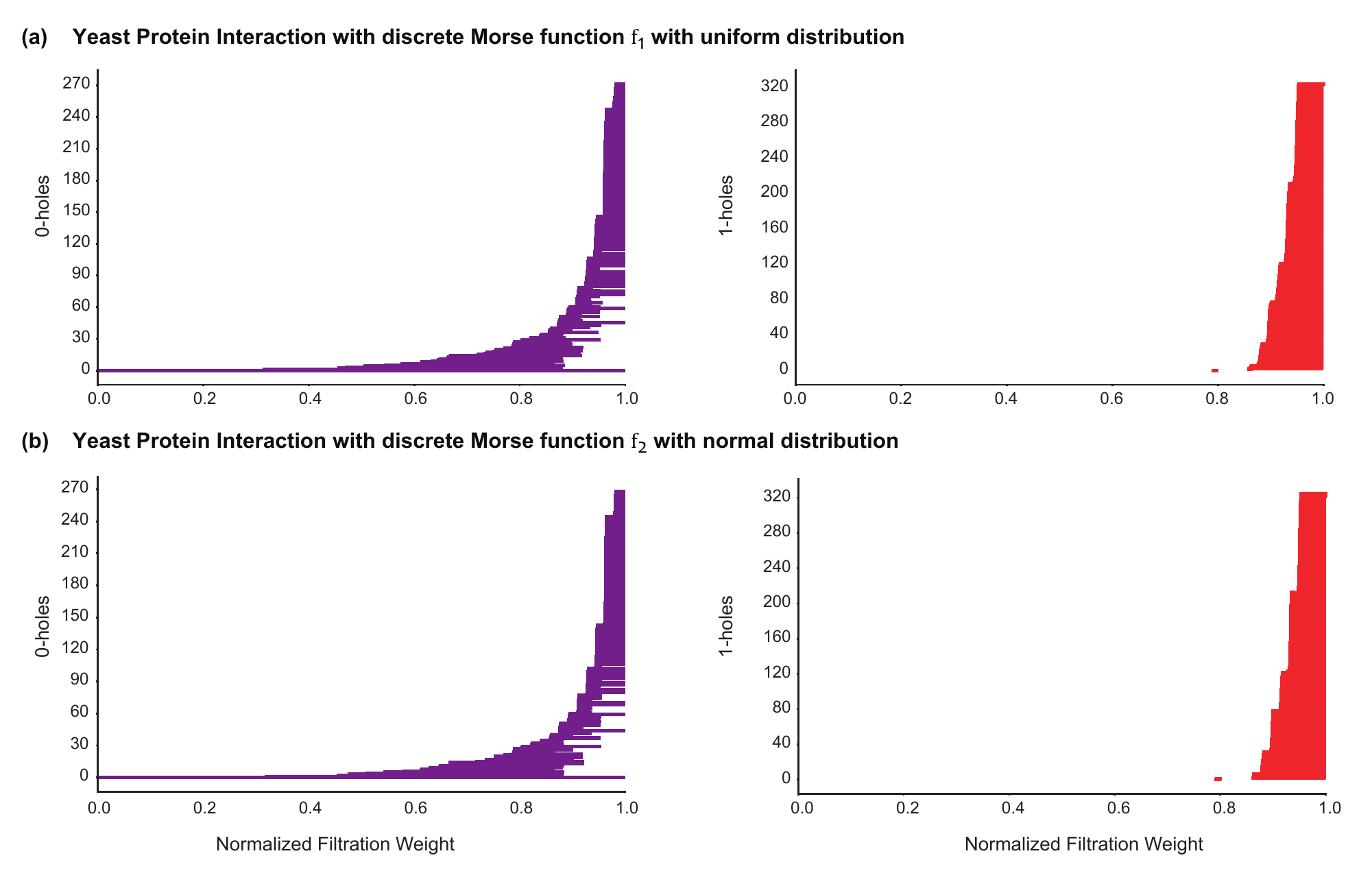}
\caption{Barcode diagrams for $H_0$ and $H_1$ in the Yeast protein interaction network computed using two discrete Morse functions $f_1$ and $f_2$ generated by making small variations in Algorithm 1. $f_1$ uses a uniform distribution on $[0.1, 0.4)$ for generating the values of $\epsilon$ in Algorithm 1, $f_2$ uses a normal distribution with mean $\mu= 0.25$ and standard deviation $\sigma=0.05$ for generating the values of $\epsilon$ in Algorithm 1.}
\end{figure*}

\subsection*{Illustration of the stability of persistence diagrams using a real network example}

As an illustration of the stability property, we compute the bottleneck distance between persistence diagrams of two discrete Morse functions $f_1$ and $f_2$ on the Yeast protein interaction network \cite{Jeong2001} generated by making small variations in Algorithm 1. The construction of $f_1$ uses a uniform distribution on $[0.1, 0.4)$ for generating the values of $\epsilon$ in Algorithm 1, while the construction of $f_2$ uses a normal distribution with mean $\mu= 0.25$ and standard deviation $\sigma=0.05$ for generating the values of $\epsilon$ in Algorithm 1. The $L^\infty$-norm of the difference $||f_1-f_2||_\infty$ is $18.3782$, while the bottleneck distance between their \textit{total} persistence diagrams (i.e., the persistence diagram of the persistence module $\maK_*:= \{\oplus_{k=0}^{\dim K} H_k(K^f(\alpha))\}_{\alpha\in \R}$), is computed to be $W_\infty(Df_1, Df_2)=0.0061$. The barcode diagrams for the $0^{th}$ and $1^{st}$ homology groups in the Yeast protein interaction network \cite{Jeong2001} with the two discrete Morse functions, $f_1$ and $f_2$, are shown in Appendix figure A1.\\


\begin{center}
\noindent{\Large \bf Supplementary Information (SI)\\ Figures}
\end{center}
\renewcommand{\thefigure}{S\arabic{figure}}
\setcounter{figure}{0}

\begin{figure}
\centering
\includegraphics[width=.7\columnwidth]{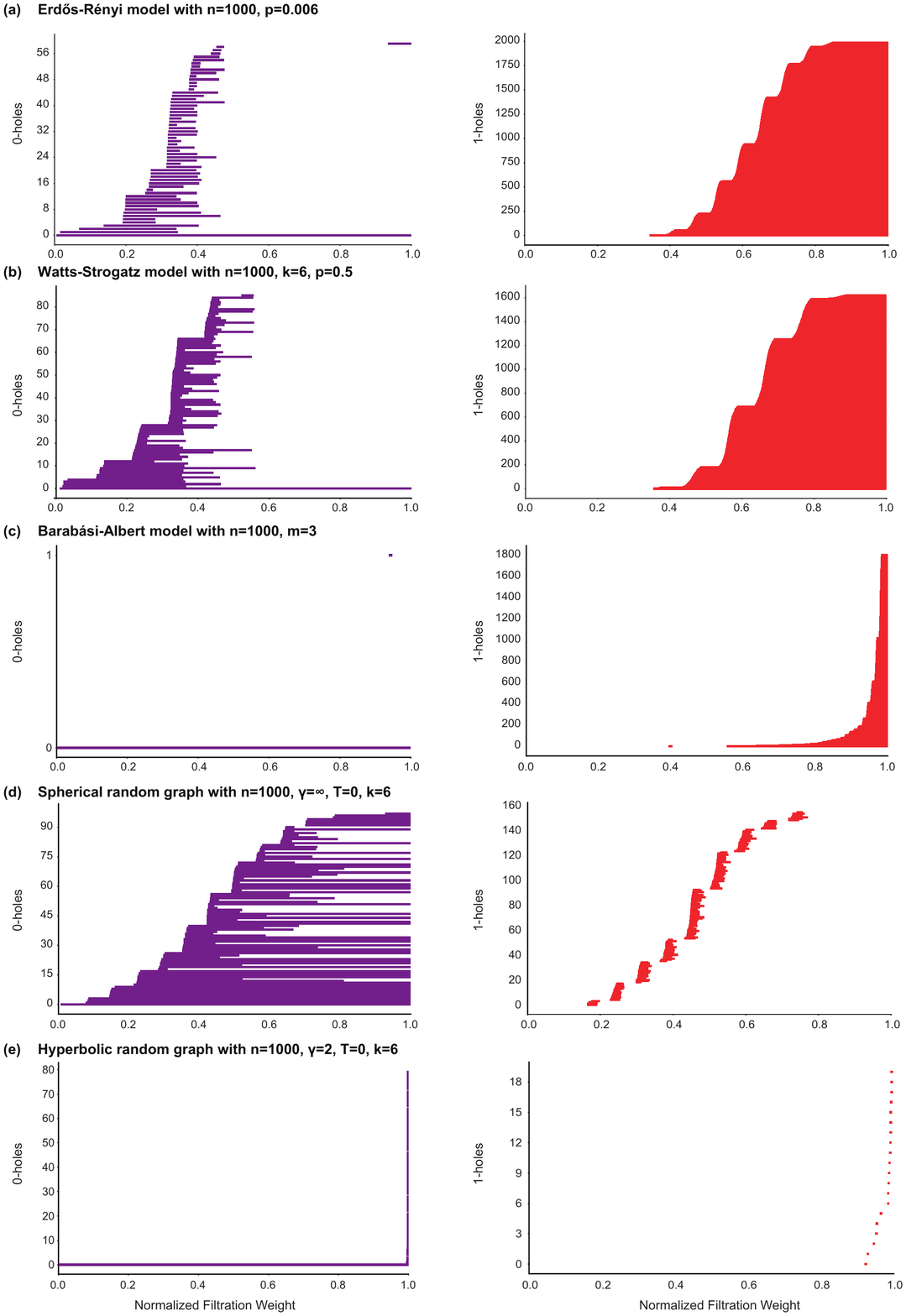}
\caption{Barcode diagrams for $H_0$ and $H_1$ in model networks with expected average degree 6.}
\end{figure}

\begin{figure}
\centering
\includegraphics[width=.7\columnwidth]{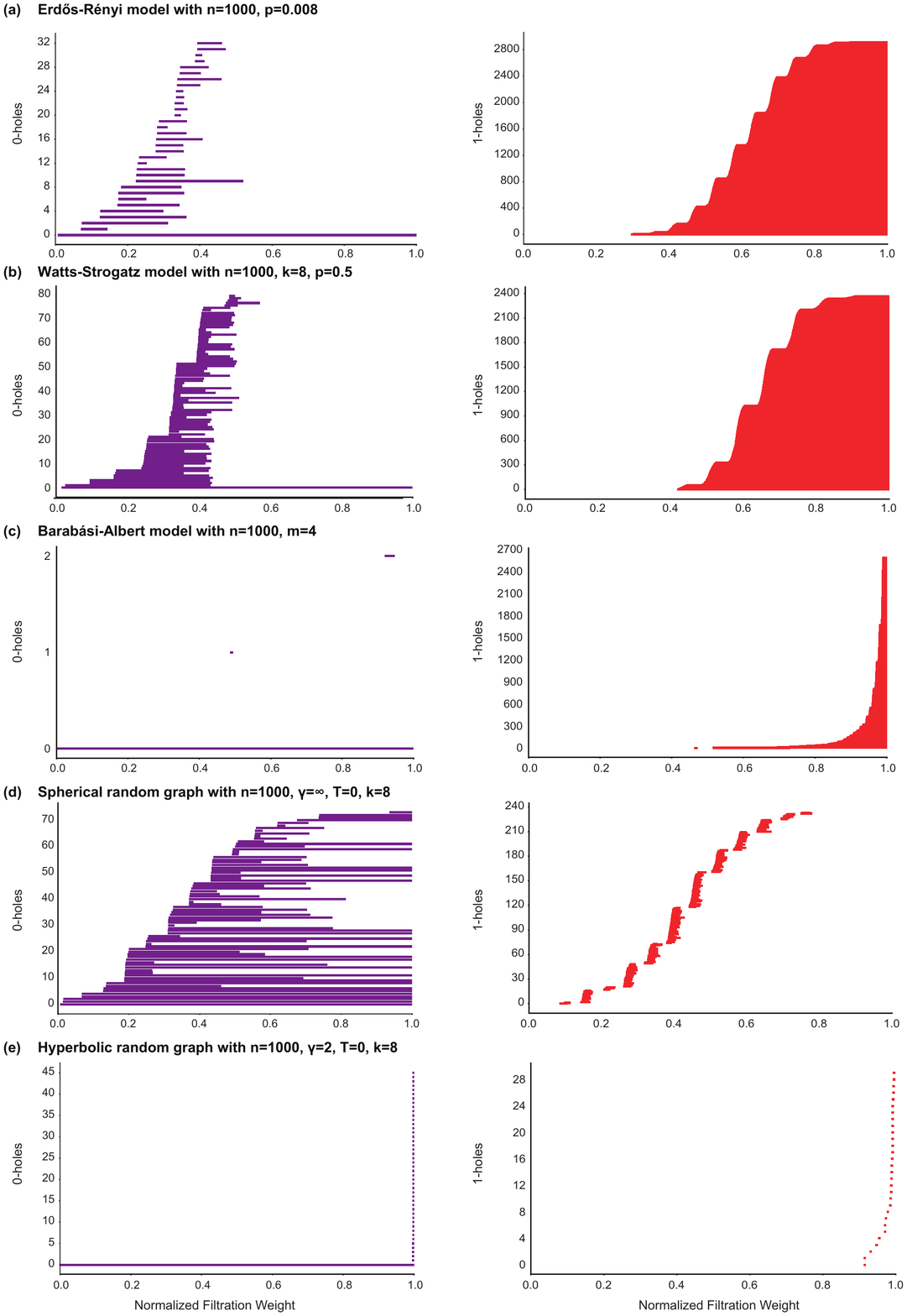}
\caption{Barcode diagrams for $H_0$ and $H_1$ in model networks with expected average degree 8.}
\end{figure}

\begin{figure}
\centering
\includegraphics[width=.7\columnwidth]{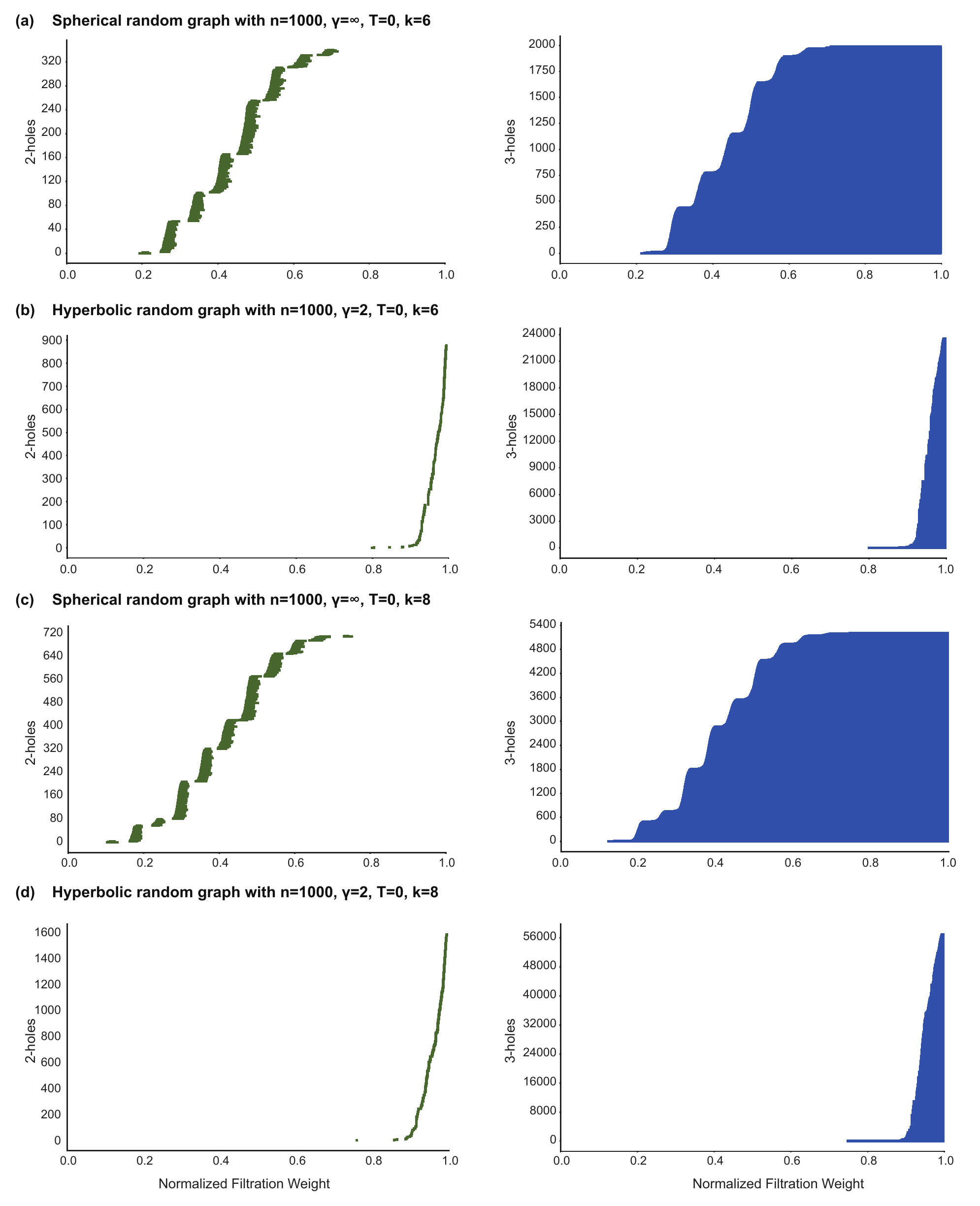}
\caption{Barcode diagrams for $H_2$ and $H_3$ in model networks with expected average degree 6 and 8. (a) Spherical random graphs with expected average degree 6. (b) Hyperbolic random graphs with expected average degree 6. (c) Spherical random graphs with expected average degree 8. (d) Hyperbolic random graphs with expected average degree 8.}
\end{figure}

\begin{figure}
\centering
\includegraphics[width=.7\columnwidth]{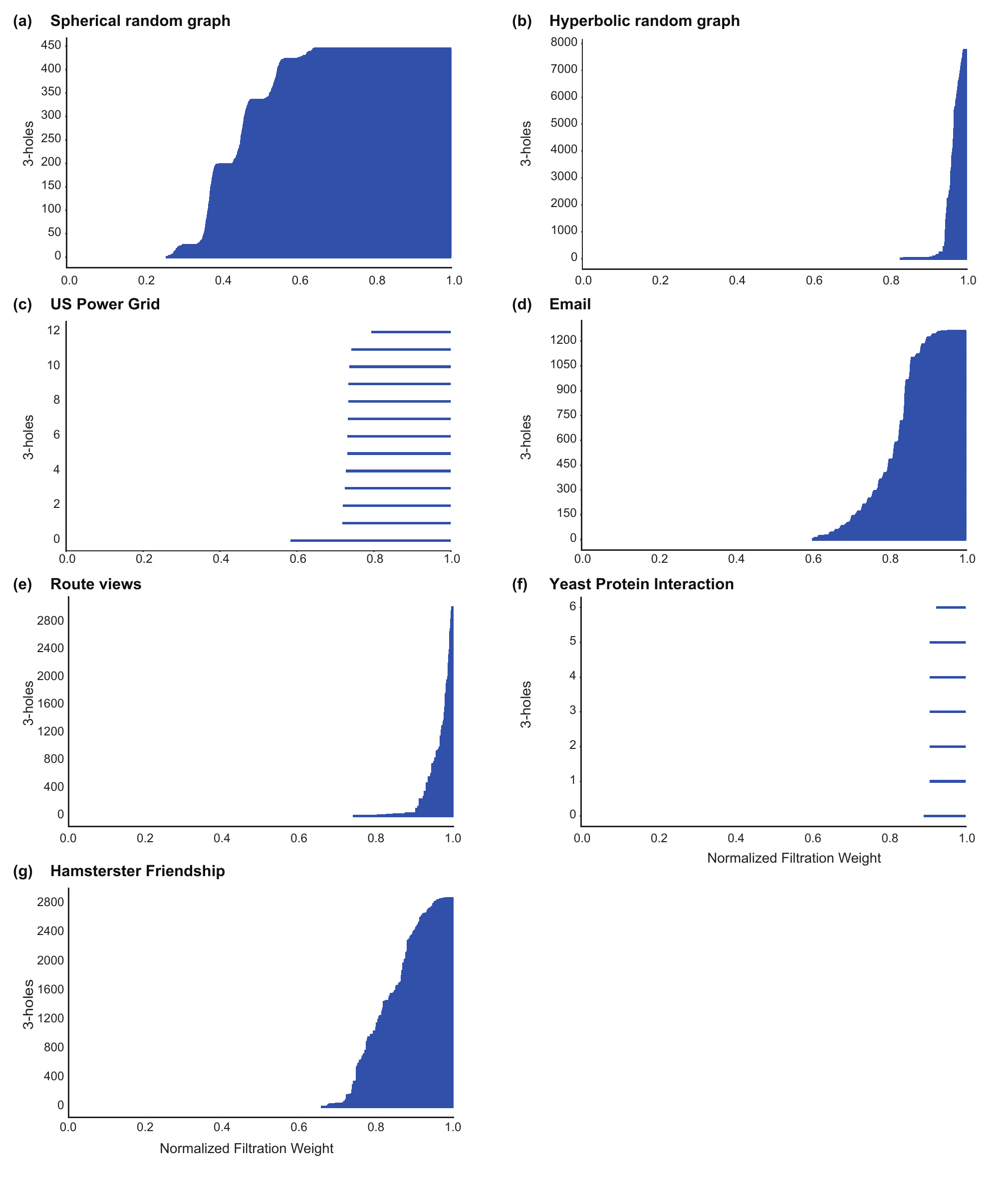}
\caption{Barcode diagrams for $H_3$ in model and real networks. (a) Spherical random graphs produced from HGG model with $n=1000$, $T=0$, $k=4$ and $\gamma=\infty$. (b) Hyperbolic random graphs produced from HGG model with $n=1000$, $T=0$, $k=4$  and $\gamma=2$. (c) US Power Grid. (d) Email communication. (e) Route views. (f) Yeast protein interaction. (g) Hamsterster friendship.}
\end{figure}

\begin{figure}
\centering
\includegraphics[width=.7\columnwidth]{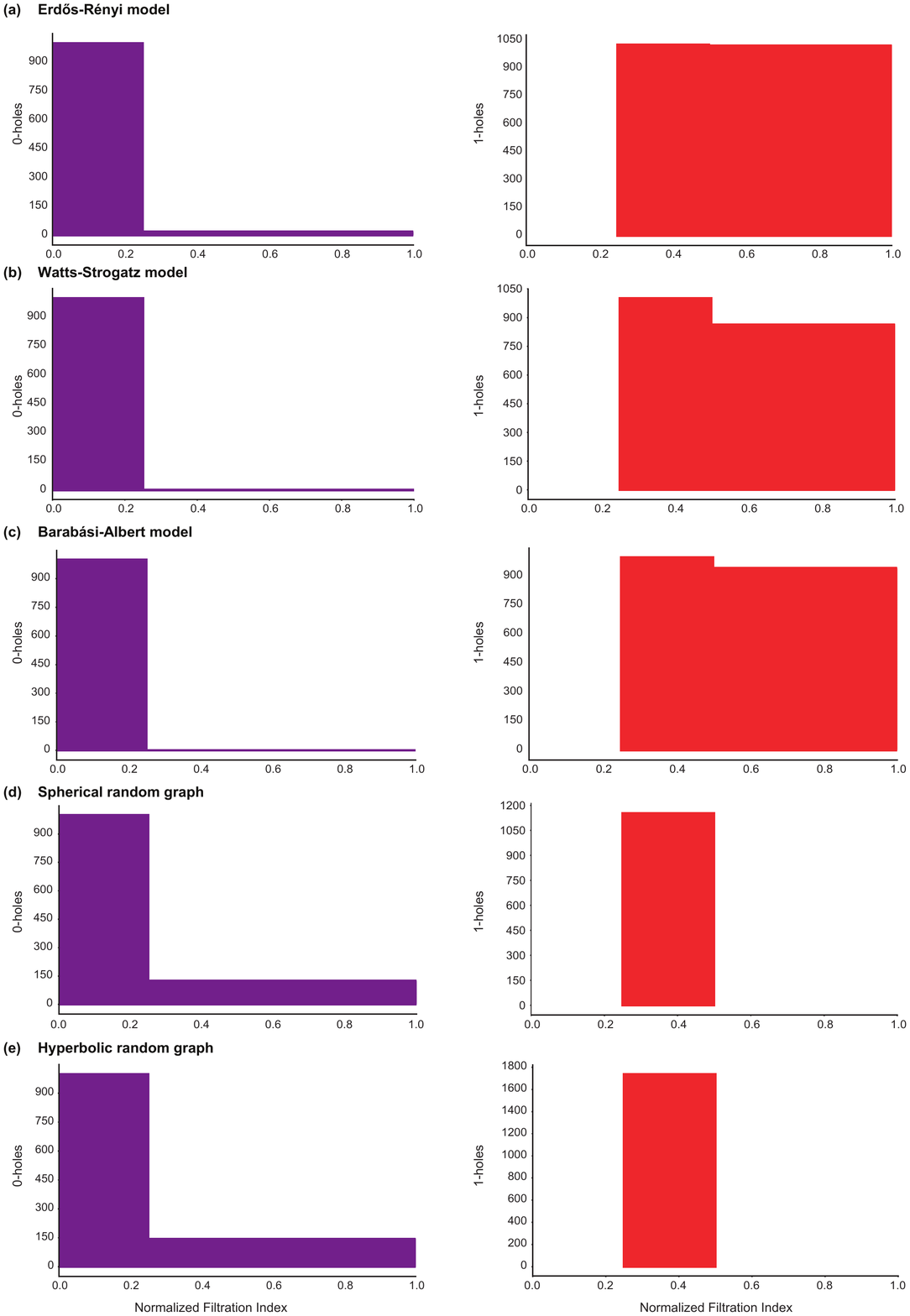}
\caption{Barcode diagrams for $H_0$ and $H_1$ in model networks with expected average degree 4 obtained using the dimensional filtration scheme used in Horak \textit{et al.} We restrict our investigation to the three-dimensional clique complex while computing these barcode diagrams for model networks using the dimensional filtration scheme of Horak \textit{et al.} In this figure, we normalize the filtration index to be in the range 0 to 1, and $p$-holes with normalized filtration index 1 indicate that they never die. (a) ER model with $n=1000$ and $p=0.004$. (b) WS model with $n=1000$, $k=4$ and $p=0.5$. (c) BA model with $n=1000$ and $m=2$. (d) Spherical random graphs produced from HGG model with $n=1000$, $T=0$, $k=4$ and $\gamma=\infty$. (e) Hyperbolic random graphs produced from HGG model with $n=1000$, $T=0$, $k=4$ and $\gamma=2$.}
\end{figure}

\begin{figure}
\centering
\includegraphics[width=.7\columnwidth]{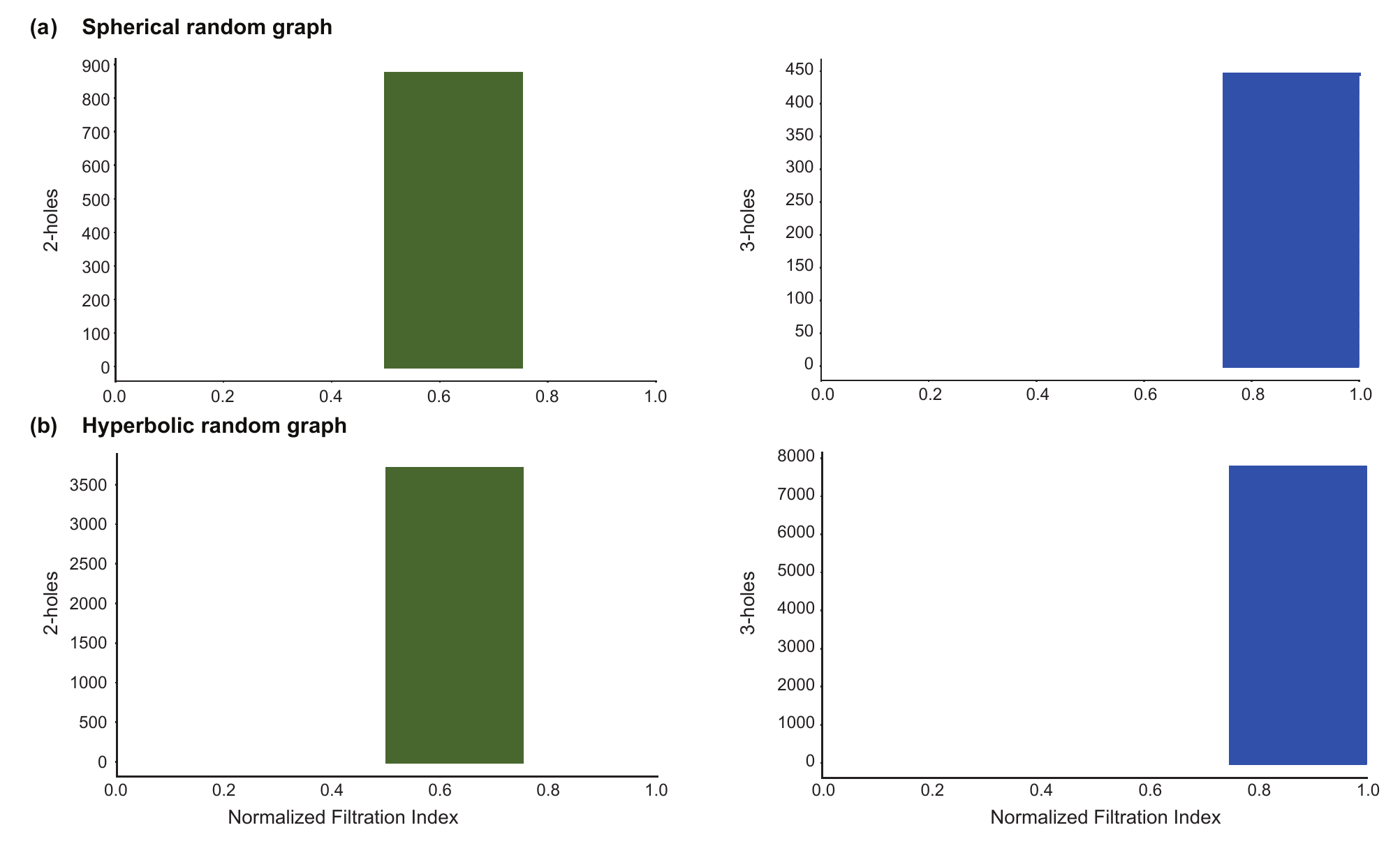}
\caption{Barcode diagrams for $H_2$ and $H_3$ in model networks with expected average degree 4 obtained using the dimensional filtration scheme used in Horak \textit{et al.} (a) Spherical random graphs produced from HGG model with $n=1000$, $T=0$, $k=4$ and $\gamma=\infty$. (b) Hyperbolic random graphs produced from HGG model with $n=1000$, $T=0$, $k=4$ and $\gamma=2$.}
\end{figure}

\begin{figure}
\centering
\includegraphics[width=.7\columnwidth]{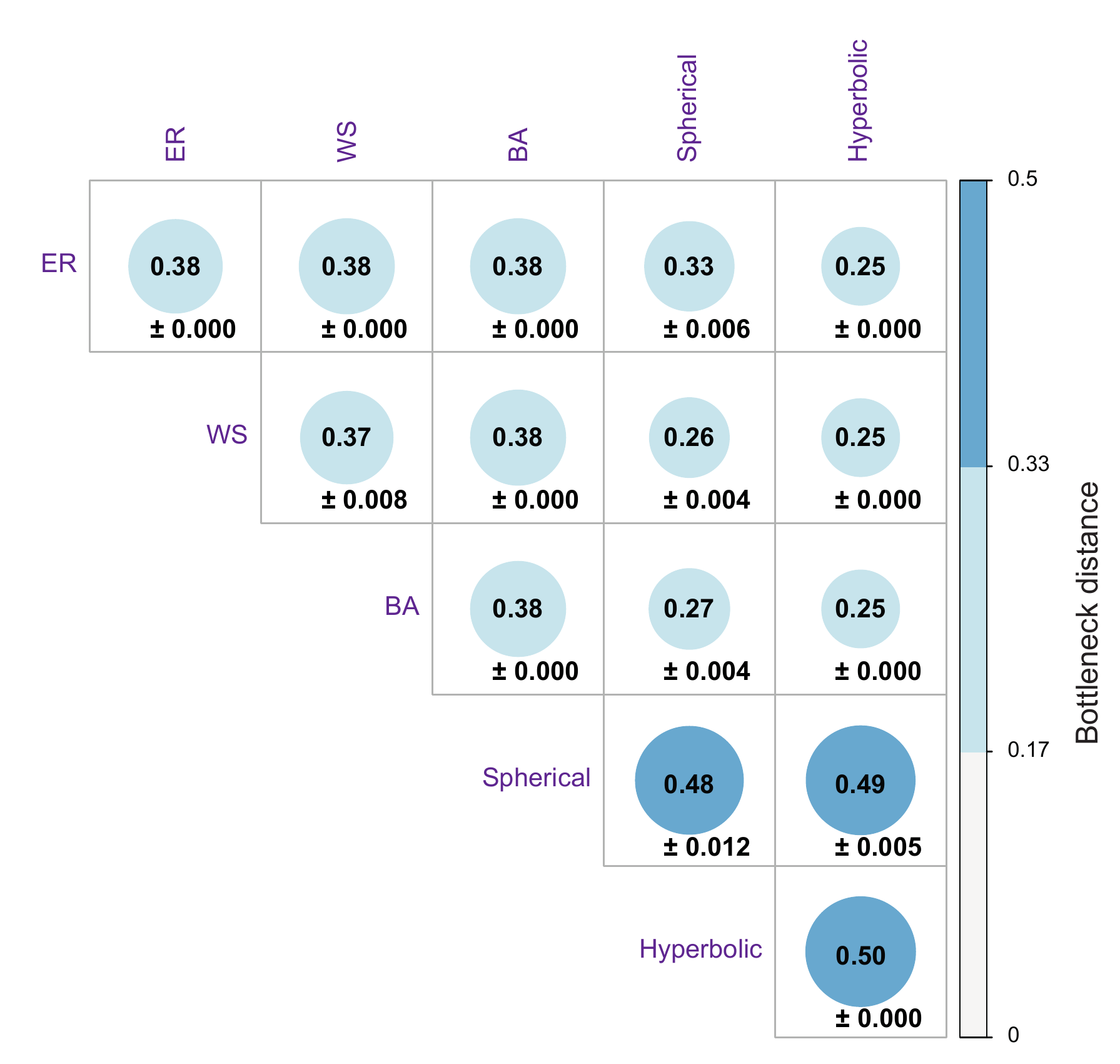}
\caption{Bottleneck distance between persistence diagrams obtained using the dimensional filtration scheme used in Horak \textit{et al.} for model networks, namely, ER model with $n=1000$ and $p=0.004$,  WS model with $n=1000$, $k=4$ and $p=0.5$, BA model with $n=1000$ and $m=2$, Spherical random graphs produced from HGG model with $n=1000$, $T=0$, $k=4$ and $\gamma=\infty$, and Hyperbolic random graphs produced from HGG model with $n=1000$, $T=0$, $k=4$ and $\gamma=2$. For each of the five model networks, 10 random samples are generated by fixing the number of vertices $n$ and other parameters of the model. We report the distance (rounded to two decimal places) between two different models as the average of the distance between each of the possible pairs of the 10 sample networks corresponding to the two models along with the standard error.}
\end{figure}


\begin{thebibliography}{10}
\expandafter\ifx\csname url\endcsname\relax
  \def\url#1{\texttt{#1}}\fi
\expandafter\ifx\csname urlprefix\endcsname\relax\def\urlprefix{URL }\fi
\expandafter\ifx\csname doiprefix\endcsname\relax\def\doiprefix{DOI }\fi
\providecommand{\bibinfo}[2]{#2}
\providecommand{\eprint}[2][]{\url{#2}}

\bibitem{Carlsson2009}
\bibinfo{author}{Carlsson, G.}
\newblock \bibinfo{journal}{\bibinfo{title}{Topology and data}}.
\newblock {\emph{ {Bulletin of the American Mathematical Society}}}
  \textbf{\bibinfo{volume}{46}}, \bibinfo{pages}{255--308}
  (\bibinfo{year}{2009}).

\bibitem{Pranav2016}
\bibinfo{author}{Pranav, P.} \emph{et~al.}
\newblock \bibinfo{journal}{\bibinfo{title}{The topology of the cosmic web in
  terms of persistent betti numbers}}.
\newblock {\emph{ {Monthly Notices of the Royal Astronomical
  Society}}} \textbf{\bibinfo{volume}{465}}, \bibinfo{pages}{4281--4310}
  (\bibinfo{year}{2016}).

\bibitem{Gunther2011}
\bibinfo{author}{G{\"u}nther, D.}, \bibinfo{author}{Reininghaus, J.},
  \bibinfo{author}{Hotz, I.} \& \bibinfo{author}{Wagner, H.}
\newblock \bibinfo{title}{Memory-efficient computation of persistent homology
  for 3d images using discrete morse theory}.
\newblock In \emph{\bibinfo{booktitle}{2011 24th SIBGRAPI Conference on
  Graphics, Patterns and Images}}, \bibinfo{pages}{25--32}
  (\bibinfo{organization}{IEEE}, \bibinfo{year}{2011}).

\bibitem{Nicolau2011}
\bibinfo{author}{Nicolau, M.}, \bibinfo{author}{Levine, A.} \&
  \bibinfo{author}{Carlsson, G.}
\newblock \bibinfo{journal}{\bibinfo{title}{Topology based data analysis
  identifies a subgroup of breast cancers with a unique mutational profile and
  excellent survival}}.
\newblock {\emph{ {Proceedings of the National Academy of Sciences
  USA}}} \textbf{\bibinfo{volume}{108}}, \bibinfo{pages}{7265--7270}
  (\bibinfo{year}{2011}).

\bibitem{Morse1934}
\bibinfo{author}{Morse, M.}
\newblock \emph{\bibinfo{title}{The calculus of variations in the large}},
  vol.~\bibinfo{volume}{18} (\bibinfo{publisher}{American Mathematical
  Society}, \bibinfo{year}{1934}).

\bibitem{Edelsbrunner2008}
\bibinfo{author}{Edelsbrunner, H.} \& \bibinfo{author}{Harer, J.}
\newblock \bibinfo{journal}{\bibinfo{title}{Persistent homology-a survey}}.
\newblock {\emph{ {Contemporary Mathematics}}}
  \textbf{\bibinfo{volume}{453}}, \bibinfo{pages}{257--282}
  (\bibinfo{year}{2008}).

\bibitem{Forman1995}
\bibinfo{author}{Forman, R.}
\newblock \bibinfo{title}{A discrete morse theory for cell complexes}.
\newblock In \bibinfo{editor}{Yau, S.-T.} (ed.)
  \emph{\bibinfo{booktitle}{Geometry, Topology and Physics for Raoul Bott}}
  (\bibinfo{publisher}{International Press of Boston}, \bibinfo{year}{1995}).

\bibitem{Forman1998}
\bibinfo{author}{Forman, R.}
\newblock \bibinfo{journal}{\bibinfo{title}{Morse theory for cell complexes}}.
\newblock {\emph{ {Advances in Mathematics}}}
  \textbf{\bibinfo{volume}{134}}, \bibinfo{pages}{90--145}
  (\bibinfo{year}{1998}).

\bibitem{Forman2002}
\bibinfo{author}{Forman, R.}
\newblock \bibinfo{journal}{\bibinfo{title}{A user's guide to discrete morse
  theory}}.
\newblock {\emph{ {S{\'e}m. Lothar. Combin.}}}
  \textbf{\bibinfo{volume}{48}}, \bibinfo{pages}{1--35} (\bibinfo{year}{2002}).

\bibitem{Watts1998}
\bibinfo{author}{Watts, D.~J.} \& \bibinfo{author}{Strogatz, S.~H.}
\newblock \bibinfo{journal}{\bibinfo{title}{Collective dynamics of small-world
  networks}}.
\newblock {\emph{ {Nature}}} \textbf{\bibinfo{volume}{393}},
  \bibinfo{pages}{440--442} (\bibinfo{year}{1998}).

\bibitem{Barabasi1999}
\bibinfo{author}{Barab{\'a}si, A.~L.} \& \bibinfo{author}{Albert, R.}
\newblock \bibinfo{journal}{\bibinfo{title}{Emergence of scaling in random
  networks}}.
\newblock {\emph{ {Science}}} \textbf{\bibinfo{volume}{286}},
  \bibinfo{pages}{509--512} (\bibinfo{year}{1999}).

\bibitem{Albert2002}
\bibinfo{author}{Albert, R.} \& \bibinfo{author}{Barab{\'a}si, A.~L.}
\newblock \bibinfo{journal}{\bibinfo{title}{Statistical mechanics of complex
  networks}}.
\newblock {\emph{ {Reviews of Modern Physics}}}
  \textbf{\bibinfo{volume}{74}}, \bibinfo{pages}{47--97}
  (\bibinfo{year}{2002}).

\bibitem{Newman2010}
\bibinfo{author}{Newman, M. E.~J.}
\newblock \emph{\bibinfo{title}{Networks: {A}n {I}ntroduction}}
  (\bibinfo{publisher}{Oxford University Press}, \bibinfo{year}{2010}).

\bibitem{Bianconi2015}
\bibinfo{author}{Bianconi, G.}
\newblock \bibinfo{journal}{\bibinfo{title}{Interdisciplinary and physics
  challenges of network theory}}.
\newblock {\emph{ {Europhysics Letters}}}
  \textbf{\bibinfo{volume}{111}}, \bibinfo{pages}{56001}
  (\bibinfo{year}{2015}).

\bibitem{Kartun-Giles2019}
\bibinfo{author}{Kartun-Giles, A.P.} \& \bibinfo{author}{Bianconi, G.}
\newblock \bibinfo{journal}{\bibinfo{title}{Beyond the clustering coefficient:
A topological analysis of node neighbourhoods in complex networks}}.
\newblock {\emph{ {Chaos, Solitons and Fractals: X}}}
  \textbf{\bibinfo{volume}{1(1)}}, \bibinfo{pages}{100004}
  (\bibinfo{year}{2019}).

\bibitem{Iacopini2019}
\bibinfo{author}{Iacopini, I.}, \bibinfo{author}{Petri, G.},
\bibinfo{author}{Barrat, A.} \& \bibinfo{author}{Latora, V.}
\newblock \bibinfo{journal}{\bibinfo{title}{Simplicial models of social
contagion}}.
\newblock {\emph{ {Nature Communications}}}
  \textbf{\bibinfo{volume}{10(1)}}, \bibinfo{pages}{2485}
  (\bibinfo{year}{2019}).

\bibitem{Ritchie2017}
\bibinfo{author}{Ritchie, M.}, \bibinfo{author}{Berthouze, L.},
\& \bibinfo{author}{Kiss, I.}
\newblock \bibinfo{journal}{\bibinfo{title}{Generation and analysis of
networks with a prescribed degree sequence and subgraph family:
higher-order structure matters}}.
\newblock {\emph{ {Journal of Complex Networks}}}
  \textbf{\bibinfo{volume}{5(1)}}, \bibinfo{pages}{1--31}
  (\bibinfo{year}{2017}).

\bibitem{De2007}
\bibinfo{author}{De~Silva, V.} \& \bibinfo{author}{Ghrist, R.}
\newblock \bibinfo{journal}{\bibinfo{title}{Homological sensor networks}}.
\newblock {\emph{ {Notices of the {A}merican mathematical
  society}}} \textbf{\bibinfo{volume}{54}} (\bibinfo{year}{2007}).

\bibitem{Horak2009}
\bibinfo{author}{Horak, D.}, \bibinfo{author}{Maleti{\'c}, S.} \&
  \bibinfo{author}{Rajkovi{\'c}, M.}
\newblock \bibinfo{journal}{\bibinfo{title}{Persistent homology of complex
  networks}}.
\newblock {\emph{ {Journal of Statistical Mechanics: Theory and
  Experiment}}} \bibinfo{pages}{P03034} (\bibinfo{year}{2009}).

\bibitem{Petri2013}
\bibinfo{author}{Petri, G.}, \bibinfo{author}{Scolamiero, M.},
  \bibinfo{author}{Donato, I.} \& \bibinfo{author}{Vaccarino, F.}
\newblock \bibinfo{journal}{\bibinfo{title}{Topological strata of weighted
  complex networks}}.
\newblock {\emph{ {PloS One}}} \textbf{\bibinfo{volume}{8}},
  \bibinfo{pages}{e66506} (\bibinfo{year}{2013}).

\bibitem{Petri2014}
\bibinfo{author}{Petri, G.} \emph{et~al.}
\newblock \bibinfo{journal}{\bibinfo{title}{Homological scaffolds of brain
  functional networks}}.
\newblock {\emph{ {Journal of The Royal Society Interface}}}
  \textbf{\bibinfo{volume}{11}}, \bibinfo{pages}{20140873}
  (\bibinfo{year}{2014}).

\bibitem{Wu2015}
\bibinfo{author}{Wu, Z.}, \bibinfo{author}{Menichetti, G.},
  \bibinfo{author}{Rahmede, C.} \& \bibinfo{author}{Bianconi, G.}
\newblock \bibinfo{journal}{\bibinfo{title}{Emergent complex network
  geometry}}.
\newblock {\emph{ {Scientific Reports}}}
  \textbf{\bibinfo{volume}{5}}, \bibinfo{pages}{10073} (\bibinfo{year}{2015}).

\bibitem{Sizemore2016}
\bibinfo{author}{Sizemore, A.}, \bibinfo{author}{Giusti, C.} \&
  \bibinfo{author}{Bassett, D.}
\newblock \bibinfo{journal}{\bibinfo{title}{Classification of weighted networks
  through mesoscale homological features}}.
\newblock {\emph{ {Journal of Complex Networks}}}
  \textbf{\bibinfo{volume}{5}}, \bibinfo{pages}{245--273}
  (\bibinfo{year}{2016}).

\bibitem{Courtney2017}
\bibinfo{author}{Courtney, O.} \& \bibinfo{author}{Bianconi, G.}
\newblock \bibinfo{journal}{\bibinfo{title}{Weighted growing simplicial
  complexes}}.
\newblock {\emph{ {Physical Review E}}}
  \textbf{\bibinfo{volume}{95}}, \bibinfo{pages}{062301}
  (\bibinfo{year}{2017}).

\bibitem{Courtney2018}
\bibinfo{author}{Courtney, O.} \& \bibinfo{author}{Bianconi, G.}
\newblock \bibinfo{journal}{\bibinfo{title}{Dense power-law networks and
  simplicial complexes}}.
\newblock {\emph{ {Physical Review E}}}
  \textbf{\bibinfo{volume}{97}}, \bibinfo{pages}{052303}
  (\bibinfo{year}{2018}).

\bibitem{Lee2012}
\bibinfo{author}{Lee, H.}, \bibinfo{author}{Kang, H.}, \bibinfo{author}{Chung,
  M.}, \bibinfo{author}{Kim, B.-N.} \& \bibinfo{author}{Lee, D.}
\newblock \bibinfo{journal}{\bibinfo{title}{Persistent brain network homology
  from the perspective of dendrogram}}.
\newblock {\emph{ {IEEE transactions on medical imaging}}}
  \textbf{\bibinfo{volume}{31}}, \bibinfo{pages}{2267--2277}
  (\bibinfo{year}{2012}).

\bibitem{Freeman1977}
\bibinfo{author}{Freeman, L.~C.}
\newblock \bibinfo{journal}{\bibinfo{title}{A set of measures of centrality
  based on betweenness}}.
\newblock {\emph{ {Sociometry}}} \textbf{\bibinfo{volume}{40}},
  \bibinfo{pages}{35--41} (\bibinfo{year}{1977}).

\bibitem{Girvan2002}
\bibinfo{author}{Girvan, M.} \& \bibinfo{author}{Newman, M.}
\newblock \bibinfo{journal}{\bibinfo{title}{Community structure in social and
  biological networks}}.
\newblock {\emph{ {Proceedings of the National Academy of Sciences
  USA}}} \textbf{\bibinfo{volume}{99}}, \bibinfo{pages}{7821--7826}
  (\bibinfo{year}{2002}).

\bibitem{Sreejith2016}
\bibinfo{author}{Sreejith, R.P.}, \bibinfo{author}{Mohanraj, K.},
  \bibinfo{author}{Jost, J.}, \bibinfo{author}{Saucan, E.} \&
  \bibinfo{author}{Samal, A.}
\newblock \bibinfo{journal}{\bibinfo{title}{Forman curvature for complex
  networks}}.
\newblock {\emph{ {Journal of Statistical Mechanics: Theory and
  Experiment}}} \bibinfo{pages}{P063206} (\bibinfo{year}{2016}).

\bibitem{Samal2018}
\bibinfo{author}{Samal, A.} \emph{et~al.}
\newblock \bibinfo{journal}{\bibinfo{title}{Comparative analysis of two
  discretizations of ricci curvature for complex networks}}.
\newblock {\emph{ {Scientific Reports}}}
  \textbf{\bibinfo{volume}{8}}, \bibinfo{pages}{8650} (\bibinfo{year}{2018}).

\bibitem{Bubenik2009}
\bibinfo{author}{Bubenik, P.}, \bibinfo{author}{Carlsson, G.},
  \bibinfo{author}{Kim, P.} \& \bibinfo{author}{Luo, Z.}
\newblock \bibinfo{journal}{\bibinfo{title}{Statistical topology via morse
  theory persistence and nonparametric estimation}}.
\newblock {\emph{ {Algebraic methods in statistics and probability
  II}}} \textbf{\bibinfo{volume}{516}}, \bibinfo{pages}{75--92}
  (\bibinfo{year}{2010}).

\bibitem{Mischaikow2013}
\bibinfo{author}{Mischaikow, K.} \& \bibinfo{author}{Nanda, V.}
\newblock \bibinfo{journal}{\bibinfo{title}{Morse theory for filtrations and
efficient computation of persistent homology}}.
\newblock {\emph{ {Discrete \& Computational Geometry}}}
\textbf{\bibinfo{volume}{50(2)}}, \bibinfo{pages}{330--353}
  (\bibinfo{year}{2013}).

\bibitem{Delgado2014}
\bibinfo{author}{Delgado-Friedrichs, O.}, \bibinfo{author}{Robins, V.} \&
  \bibinfo{author}{Sheppard, A.}
\newblock \bibinfo{title}{Morse theory and persistent homology for topological
  analysis of 3d images of complex materials}.
\newblock In \emph{\bibinfo{booktitle}{2014 IEEE International Conference on
Image Processing (ICIP)}}, \bibinfo{pages}{4872--4876}
  (\bibinfo{organization}{IEEE}, \bibinfo{year}{2014}).

\bibitem{Bollobas1998}
\bibinfo{author}{Bollobas, B.}
\newblock \emph{\bibinfo{title}{Modern Graph Theory}}
  (\bibinfo{publisher}{Springer}, \bibinfo{year}{1998}).

\bibitem{Zomorodian2005}
\bibinfo{author}{Zomorodian, A.} \& \bibinfo{author}{Carlsson, G.}
\newblock \bibinfo{journal}{\bibinfo{title}{Computing persistent homology}}.
\newblock {\emph{ {Discrete {\&} Computational Geometry}}}
  \textbf{\bibinfo{volume}{33}}, \bibinfo{pages}{249--274}
  (\bibinfo{year}{2005}).

\bibitem{Munkres2018}
\bibinfo{author}{Munkres, J.}
\newblock \emph{\bibinfo{title}{Elements of algebraic topology}}
  (\bibinfo{publisher}{CRC Press}, \bibinfo{year}{2018}).

\bibitem{Cohen-Steiner2007}
\bibinfo{author}{Cohen-Steiner, D.}, \bibinfo{author}{Edelsbrunner, H.} \&
  \bibinfo{author}{Harer, J.}
\newblock \bibinfo{journal}{\bibinfo{title}{Stability of persistence
  diagrams}}.
\newblock {\emph{ {Discrete \& Computational Geometry}}}
  \textbf{\bibinfo{volume}{37}}, \bibinfo{pages}{103--120}
  (\bibinfo{year}{2007}).

\bibitem{Kerber2017}
\bibinfo{author}{Kerber, M.}, \bibinfo{author}{Morozov, D.} \&
  \bibinfo{author}{Nigmetov, A.}
\newblock \bibinfo{journal}{\bibinfo{title}{Geometry helps to compare
  persistence diagrams}}.
\newblock {\emph{ {J. Exp. Algorithmics}}}
  \textbf{\bibinfo{volume}{22}}, \bibinfo{pages}{1.4:1--1.4:20}
  (\bibinfo{year}{2017}).

\bibitem{Chazal2008}
\bibinfo{author}{Chazal, F.}, \bibinfo{author}{Cohen-Steiner, D.},
\bibinfo{author}{Guibas, L. J.} \&  \bibinfo{author}{Oudot, S.},
\newblock \bibinfo{journal}{\bibinfo{title}{Stability of persistence
  diagrams revisited}},
\newblock {{INRIA Research report RR-6568 available at:}
  \url{https://hal.inria.fr/inria-00292566v1/}} (\bibinfo{year}{2008}).

\bibitem{Erdos1961}
\bibinfo{author}{Erd\"{o}s, P.} \& \bibinfo{author}{R{\'e}nyi, A.}
\newblock \bibinfo{journal}{\bibinfo{title}{On the evolution of random
  graphs}}.
\newblock {\emph{ {Bull. Inst. Internat. Statist}}}
  \textbf{\bibinfo{volume}{38}}, \bibinfo{pages}{343--347}
  (\bibinfo{year}{1961}).

\bibitem{Krioukov2010}
\bibinfo{author}{Krioukov, D.}, \bibinfo{author}{Papadopoulos, F.},
  \bibinfo{author}{Kitsak, M.}, \bibinfo{author}{Vahdat, A.} \&
  \bibinfo{author}{Bogun{\'a}, M.}
\newblock \bibinfo{journal}{\bibinfo{title}{Hyperbolic geometry of complex
  networks}}.
\newblock {\emph{ {Physical Review E}}}
  \textbf{\bibinfo{volume}{82}}, \bibinfo{pages}{036106}
  (\bibinfo{year}{2010}).

\bibitem{Aldecoa2015}
\bibinfo{author}{Aldecoa, R.}, \bibinfo{author}{Orsini, C.} \&
  \bibinfo{author}{Krioukov, D.}
\newblock \bibinfo{journal}{\bibinfo{title}{Hyperbolic graph generator}}.
\newblock {\emph{ {Computer Physics Communications}}}
  \textbf{\bibinfo{volume}{196}}, \bibinfo{pages}{492--496}
  (\bibinfo{year}{2015}).

\bibitem{Jeong2001}
\bibinfo{author}{Jeong, H.}, \bibinfo{author}{Mason, S.~P.},
  \bibinfo{author}{Barab{\'a}si, A.~L.} \& \bibinfo{author}{Oltvai, Z.~N.}
\newblock \bibinfo{journal}{\bibinfo{title}{Lethality and centrality in protein
  networks}}.
\newblock {\emph{ {Nature}}} \textbf{\bibinfo{volume}{411}},
  \bibinfo{pages}{41--42} (\bibinfo{year}{2001}).

\bibitem{Rual2005}
\bibinfo{author}{Rual, J.~F.} \emph{et~al.}
\newblock \bibinfo{journal}{\bibinfo{title}{Towards a proteome-scale map of the
  human protein--protein interaction network}}.
\newblock {\emph{ {Nature}}} \textbf{\bibinfo{volume}{437}},
  \bibinfo{pages}{1173--1178} (\bibinfo{year}{2005}).

\bibitem{Leskovec2007}
\bibinfo{author}{Leskovec, J.}, \bibinfo{author}{Kleinberg, J.} \&
  \bibinfo{author}{Faloutsos, C.}
\newblock \bibinfo{journal}{\bibinfo{title}{Graph evolution: Densification and
  shrinking diameters}}.
\newblock {\emph{ {ACM Transactions on Knowledge Discovery from
  Data (TKDD)}}} \textbf{\bibinfo{volume}{1}}, \bibinfo{pages}{2}
  (\bibinfo{year}{2007}).

\bibitem{Subelj2011}
\bibinfo{author}{{\v{S}}ubelj, L.} \& \bibinfo{author}{Bajec, M.}
\newblock \bibinfo{journal}{\bibinfo{title}{Robust network community detection
  using balanced propagation}}.
\newblock {\emph{ {European Physical Journal B}}}
  \textbf{\bibinfo{volume}{81}}, \bibinfo{pages}{353--362}
  (\bibinfo{year}{2011}).

\bibitem{Guimera2003}
\bibinfo{author}{Guimera, R.}, \bibinfo{author}{Danon, L.},
  \bibinfo{author}{Diaz-Guilera, A.}, \bibinfo{author}{Giralt, F.} \&
  \bibinfo{author}{Arenas, A.}
\newblock \bibinfo{journal}{\bibinfo{title}{Self-similar community structure in
  a network of human interactions}}.
\newblock {\emph{ {Physical Review E}}}
  \textbf{\bibinfo{volume}{68}}, \bibinfo{pages}{065103}
  (\bibinfo{year}{2003}).

\bibitem{Kunegis2013}
\bibinfo{author}{Kunegis, J.}
\newblock \bibinfo{title}{Konect: the {K}oblenz network collection}.
\newblock In \emph{\bibinfo{booktitle}{Proceedings of the 22nd International
  Conference on World Wide Web companion}}, \bibinfo{pages}{1343--1350}
  (\bibinfo{publisher}{ACM}, \bibinfo{address}{New York, NY, USA},
  \bibinfo{year}{2013}).

\bibitem{Lewiner2003}
\bibinfo{author}{Lewiner, T.}, \bibinfo{author}{Lopes, H.} \&
  \bibinfo{author}{Tavares, G.}
\newblock \bibinfo{journal}{\bibinfo{title}{Toward optimality in discrete morse
  theory}}.
\newblock {\emph{ {Experimental Mathematics}}}
  \textbf{\bibinfo{volume}{12}}, \bibinfo{pages}{271--285}
  (\bibinfo{year}{2003}).

\bibitem{Maria2014}
\bibinfo{author}{Maria, C.}, \bibinfo{author}{Boissonnat, J.-D.},
  \bibinfo{author}{Glisse, M.} \& \bibinfo{author}{Yvinec, M.}
\newblock \bibinfo{title}{The {GUDHI} library: Simplicial complexes and
  persistent homology}.
\newblock In \emph{\bibinfo{booktitle}{International Congress on Mathematical
  Software}}, \bibinfo{pages}{167--174} (\bibinfo{organization}{Springer},
  \bibinfo{year}{2014}).

\bibitem{Ghrist2008}
\bibinfo{author}{Ghrist, R.}
\newblock \bibinfo{journal}{\bibinfo{title}{Barcodes: the persistent topology
  of data}}.
\newblock {\emph{ {Bulletin of the American Mathematical Society}}}
  \textbf{\bibinfo{volume}{45}}, \bibinfo{pages}{61--75}
  (\bibinfo{year}{2008}).
  
\bibitem{Dummit2003}
\bibinfo{author}{Dummit, D.} \& \bibinfo{author}{Foote, R.}
\newblock \emph{\bibinfo{title}{Abstract algebra}} (\bibinfo{publisher}{Wiley},
  \bibinfo{year}{2003}), \bibinfo{edition}{3} edn.
  
\bibitem{Forman2005}
\bibinfo{author}{Forman, R.}
\newblock {\bibinfo{title}{Some applications of combinatorial differential topology}}.
\newblock In \bibinfo{editor}{Lyubich, M. } \& \bibinfo{editor}{Takhtajan, L.} (eds.)
\emph{\bibinfo{booktitle}{Graphs and patterns in mathematics and theoretical physics}},
\bibinfo{volume}{73}, \bibinfo{pages}{281--313} (\bibinfo{publisher}{American Mathematical
Society (AMS)}, \bibinfo{year}{2005}).
 
\bibitem{Ayala2009}
\bibinfo{author}{Ayala, R.}, \bibinfo{author}{Fern\'{a}ndez, L. M.} \& \bibinfo{author}{Vilches, J.A.},
\newblock \bibinfo{journal}{\bibinfo{title}{Characterizing equivalent discrete Morse functions,}}
\newblock {\emph{ {Bulletin of the Brazilian Mathematical Society, New Series}}}
  \textbf{\bibinfo{volume}{40 (2)}}, \bibinfo{pages}{225--235}
  (\bibinfo{year}{2009}).

\bibitem{Bauer2012}
\bibinfo{author}{Bauer, U.}, \bibinfo{author}{Lange, C.} \& \bibinfo{author}{Wardetzky, M.},
\newblock \bibinfo{journal}{\bibinfo{title}{Optimal Topological Simplification of Discrete
Functions on Surfaces,}}
\newblock {\emph{ {Discrete \& Computational Geometry}}}
  \textbf{\bibinfo{volume}{47(2)}}, \bibinfo{pages}{347--377}
  (\bibinfo{year}{2012}).

\bibitem{DiFabio2015}
\bibinfo{author}{Di Fabio, B.} \&  \bibinfo{author}{Ferri, M.}
\newblock \bibinfo{journal}{\bibinfo{title}{Comparing Persistence Diagrams Through Complex Vectors}}.
\newblock In \emph{\bibinfo{booktitle}{Image Analysis and Processing --- ICIAP 2015}},
\bibinfo{pages}{294--305}
(\bibinfo{publisher}{Springer International Publishing},
\bibinfo{address}{Cham}, \bibinfo{year}{2015}).

\end{thebibliography}
\end{document}